\documentclass[a4letter]{PoS}

\let\OLDthebibliography\thebibliography
\renewcommand\thebibliography[1]{
  \OLDthebibliography{#1} 
  \setlength{\parskip}{0pt}
  \setlength{\itemsep}{0pt plus 0.2ex}
  \let\thefootnote\relax\footnotetext{ICRC proceedings identified by number \#XXX can be found online at the PoS webpage by using the link:
\href{http://pos.sissa.it/archive/conferences/236/xxx/ICRC2015$\_$xxx.pdf}{\tt http://pos.sissa.it/archive/conferences/236/xxx/ICRC2015$\_$xxx.pdf}}

}

\newcommand\thbn{\vartheta}
\renewcommand{\deg}{^{\circ}}
\newcommand{\vsh}{v_{\rm sh}}
\newcommand{\xcr}{\xi_{\rm cr}}
\newcommand{\esh}{E_{\rm sh}}
\newcommand{\Em}{E_{\rm max}}
\newcommand{\Eme}{E_{\rm max,e}}
\newcommand{\Emp}{E_{\rm max,p}}
\newcommand{\Ei}{E_{\rm inj}}
\newcommand{\Kep}{K_{\rm ep}}
\newcommand{\Ekn}{E_{\rm knee}}
\newcommand{\Ecut}{E_{\gamma,{\rm cut}}}

\title{Cosmic-ray Acceleration and Propagation}

\ShortTitle{CR Acceleration and Propagation}

\author{\speaker{Damiano Caprioli}\\
        Princeton University\\
        4 Ivy Ln. - Princeton, NJ 08544 - USA\\
        E-mail: \email{caprioli@astro.princeton.edu}}

\abstract{
The origin of cosmic rays (CRs) has puzzled scientists since the pioneering discovery by Victor Hess in 1912. 
In the last decade, however, modern supercomputers have opened a new window on the processes regulating astrophysical collisionless plasmas, allowing the study of CR acceleration via first-principles kinetic simulations. 
At the same time, a new-generation of X-ray and $\gamma$-ray telescopes has been collecting evidence that Galactic CRs are accelerated in the blast waves of supernova remnants (SNRs).
I present state-of-the-art particle-in-cells simulations of non-relativistic shocks, in which ion and electron acceleration efficiency and magnetic field amplification are studied in detail as a function of the shock parameters. 
I then discuss the theoretical and observational counterparts of these findings, comparing them with predictions of diffusive shock acceleration theory and with multi-wavelength observations of young SNRs. 
I especially outline some major open questions, such as the possible causes of the steep CR spectra inferred from $\gamma$-ray observations of SNRs and the origin of the knee in the Galactic CR spectrum. Finally, I put such a theoretical understanding in relation with CR propagation in the Galaxy in order to bridge the gap between acceleration in sources and measurements of CRs at Earth.}

\FullConference{The 34th International Cosmic Ray Conference,\\
		30 July- 6 August, 2015\\
		The Hague, The Netherlands}

\begin{document}

\section{Introduction}
The quest for the sources of cosmic rays (CRs) has involved several generations of observers and theorists for more than a century.
The 34$^{\rm th}$ International Cosmic Ray Conference held in The Hague, in which about 1,300 contributions were presented, is just the most recent milestone in such a quest.
Unraveling the physical mechanisms responsible for the acceleration of the fastest massive particles in the universe has traditionally moved along three paths: direct detection of CR fluxes at Earth, also by means of balloons, spacecraft, and satellites; observation of non-thermal emission from astrophysical objects; and theoretical interpretation of such a wealth of data.

In the last few decades, state-of-the-art X-ray and $\gamma$-ray telescopes have opened new windows on the non-thermal universe, providing us unprecedented high-resolution images and spectra of CR candidate sources, joining the radio telescopes that already in the '50s have revealed the presence of relativistic electrons in Galactic objects such as supernova remnants (SNRs).
At the same time, the advent of modern supercomputers has allowed numerical plasma simulations to become prominent tools for studying the complex interplay between energetic particles and electromagnetic fields, which is at the basis of acceleration in collisionless plasmas.
I briefly summarize the current status of the direct detection of CRs with energies $\lesssim 10^8$\,GeV, which are likely accelerated in our Galaxy, and critically review the long-standing idea that \emph{diffusive shock acceleration} in SNR blast waves is the mechanism responsible for their acceleration (\emph{SNR paradigm}, \S\ref{sec:SNR}).
The most recent findings obtained with kinetic simulations of non-relativistic shock waves are outlined in \S\ref{sec:sims}, along with their connection to the non-thermal phenomenology of SNRs, especially $\gamma$-ray observations (\S\ref{sec:gamma}).
Finally, in \S\ref{sec:prop} I discuss the bridge between the non-thermal SNR phenomenology and the CR fluxes measured at Earth, and in particular the current uncertainties in the escape of accelerated particles from their sources and in the self-confinement of energetic particles.

\section{The (almost) universal spectrum of cosmic rays \label{sec:spectrum}}
The CR spectrum measured at Earth spans more than ten orders of magnitude in energy, from fractions of GeV up to about $10^{11}$GeV, with the remarkable regularity of a power-law with spectral index $\sim$\,3.
Below a few tens of GeV the CR spectrum is modulated by the solar wind, which has a screening effect on Galactic CRs.
Yet, the Voyager I spacecraft, which in 2013 became the first man-made object that has left the heliosphere, has directly measured the pristine interstellar CR spectrum of electrons, H, and He \cite{stone+13,potgieter13}.
Once combined with the local spectra measured by PAMELA and AMS-02, this additional information will allow to better understand solar modulation, and in turn the spectrum of low-energy Galactic CRs\footnote{An \emph{indirect} measurement comes from the $\gamma-$ray emission from molecular clouds (MCs) in the Gould Belt \cite{neronov+12}.}. 

At $\Ekn\approx 5\times 10^{6}$\,GeV (the \emph{knee}), the all-particle CR spectrum steepens from about $E^{-2.65}$ to $E^{-3.1}$ and its chemical composition becomes increasingly heavy up to $\sim 10^8$\,GeV.
This suggests a homogenous class of sources in which CR are accelerated via a rigidity-dependent mechanism up to $Z\,\Ekn$, where $Z$ is the nuclear charge, with the change in slope due to the convolution of $Z$-dependent cutoffs of different species (see, e.g., \cite{hoerandel05,ba12a, nuclei, gaisser+13}).
Recently, PAMELA \cite{PAMELA11} and AMS-02 \cite{AMS15} revealed an additional feature in the H and He spectra, i.e., a quite abrupt flattening of about 0.14 (AMS-02) and 0.22 (PAMELA) in spectral slope around 200\,GeV/nucleon.
Above 300\,GeV/nucleon these slopes  are consistent with the results of previous experiments, also for heavier nuclei (e.g., CREAM \cite{CREAM09,CREAM10}, TRACER \cite{TRACER11}, BESS \cite{BESS15}, and ATIC-2 \cite{ATIC09}).
The very presence of such a spectral break suggests a first-order correction to the ``universal'' CR slope, either at the acceleration or at the transport stage.
Also, the H spectrum is steeper by about 0.1 in spectral slope with respect to He and heavier elements (e.g., \cite{CREAM10}), which means that the dependence of acceleration and propagation on rigidity only may be questioned at this level of accuracy.
The reader can refer to \cite{serpico15} for a thorough discussion of such ``anomalies'', as well as for the implications of the peculiar spectra of positrons and antiprotons measured by PAMELA and AMS-02.

The last few years have reserved some surprises also for what concerns the nature of the knee, which is usually interpreted as due to the intrinsic maximum rigidity that particles can achieve in their sources
(another possibility being that $\Ekn$ is determined by the deterioration of Galactic confinement, e.g., \cite{giacinti+15}).
The ARGO--YBJ experiment has measured the chemical composition of CRs between $\sim 30\,$TeV and $\sim 3$ PeV \cite{ARGO15}, finding that the light (H+He) component shows a gradual change of slope above 700 TeV, a factor of $\sim\,30$ below the proton knee reported by KASCADE-Grande \cite{KASCADE13}, while confirming the canonical value for the all-particle knee of $\sim 5\times 10^6\,$GeV.

Such deviations from straight power-laws and rigidity scalings are clear examples that CR physics has entered its {\em precision era}, in which a simple, global (and hopefully elegant) paradigm for the origin and the transport of energetic particles needs to be complemented with first-order corrections to account for a richer phenomenology.


In the quest for the actual CR sources, the \emph{Hillas criterion} (\cite{hillas84}, though see \cite{cavallo78} for an earlier formulation) can rule out objects that lack the minimum magnetic field strength and system size necessary to accelerate particles up to a given energy.
CRs with rigidities $\gtrsim 10^8$\,GV have a gyroradius $r_L\simeq  E/(Z\times 10^6\,{\rm GeV})(B/\mu{\rm G})^{-1}\, {\rm pc}$ that exceeds the size of the Galactic disk, which suggests their sources to be extra-galactic objects, such as active galactic nuclei \cite{Murase+12,espresso}, $\gamma$-ray bursts \cite{waxman95,vietri95}, and newly-born millisecond pulsars \cite{Blasi+00,Fang+12}.
Addressing the nature of the transition from Galactic to extra-galactic CRs is beyond the scope of this review, but it is indeed important to check whether CRs can be accelerated at least up to $\Ekn$ in Galactic accelerators.

\subsection{The SNR paradigm\label{sec:SNR}}
\noindent{\bf Energetics.}
SN explosions were associated to CR acceleration for the first time by Baade and Zwicky in 1934 \cite{bz34}.
Their energetic argument is still valid today, even if in their pioneering paper they argued for an extra-galactic origin of all the CRs.
Assuming a rate of $\mathcal{R_{\rm SN}}\approx 1-3$ per century, Galactic SN explosions can account for the luminosity of CRs below the knee if about 5--15\% of the ejecta kinetic energy is channeled into accelerated particles (see, e.g., \cite{hillas05}).
However, the appeal of SNe as CR sources is not merely limited to an energetic argument, since in principle also stellar winds may provide an adequate energy reservoir (see \cite{bykov14} for a recent review).\\

\noindent{\bf Universal power-law spectra.}
Despite the spread in the environmental parameters intrinsic in any class of astrophysical objects, the regularity of the CR spectrum below the knee requires an acceleration mechanism returning a universal power-law spectrum, and that such a power-law nature must be preserved by propagation in the Galaxy. 
The spectral features outlined above may either arise from the ``imperfections'' of such universal models or reflect the diverse taxonomy of CR sources.
\emph{First-order Fermi acceleration} \cite{fermi49} applied to SNR blast waves has what it takes to be such a universal mechanism, as it has been put forward independently by several scientists in the late '70s \cite{krymskii77, axford+78, bell78a, bo78}.
In such a diffusive shock acceleration (DSA), particles with gyroradii larger than the shock thickness can be repeatedly scattered back and forth across the shock, gaining energy as if they were squeezed between converging walls.
Since both the energy gain per cycle and the probability of being advected away from the acceleration region are controlled by the shock hydrodynamics only, accelerated particles develop power-law distributions whose spectral index $\alpha$ is fully determined by the downstream/upstream density compression ratio, $r$.
For monoatomic gas with adiabatic index $\gamma=5/3$ and strong shocks with sonic Mach number $M_s=\vsh/c_s\gg 1$, where $\vsh$ and $c_s$ are the shock velocity and the  speed of sound, the compression ratio $r\to 4$ and the differential momentum spectrum of accelerated particles reads
\begin{equation}\label{eq:universal}
N(p)=4\pi p^2 f(p);\quad f(p)\propto p^{-\frac{3r}{r-1}} \propto p^{-4}.
\end{equation}
The energy spectrum in turn is $N(E)dE=4\pi p^2 f(p)dp$: if particles are relativistic ($E\propto p$), then $N(E)\propto E^{-2}$, while for non-relativistic particles ($E\propto p^2\to dE/dp\propto p$) one gets $N(E)\propto E^{-1.5}$. 
Since SNR shocks have $M_s\gg1$, they are expected to accelerate CRs with spectra $N(E)\propto E^{-2}$.\\

\noindent{\bf Diffusive transport in the Milky Way.}
The CR Galactic residence time can be estimated thanks to \emph{radioactive clocks} such as $^{10}$Be (only available to relatively low energies) and to the ratios of secondary to primary species such as B/C, Li/C, (Sc+V)/Fe, which return the grammage traversed by primary CRs in the Galaxy.
All of these measurements suggest that CRs with $\sim 10$\,GeV spend $\sim\,10^8$\,yr in the Galaxy before escaping, significantly longer that the ballistic propagation time.
If CRs are produced in the disk and diffusively escape at some distance $H$ ($\sim$\,a few kpc) in the halo, the Galactic residence time is $\tau_{\rm gal}(E)\approx H^2/D_{\rm gal}(E)$, where $D_{\rm gal}(E)$ is the diffusion coefficient that parametrizes CR transport in the Galaxy, assumed homogeneous and isotropic.

The energy dependence of such primary/secondary ratios scales as $\tau_{\rm gal}\propto E^{-\delta}$ and is crucial for connecting the spectra injected at sources ($N_{\rm s}\propto E^{-\alpha}$) with those measured at Earth, ($\propto E^{-2.65}$ below the knee \cite{PAMELA14,AMS15}).
The equilibrium CR spectrum can in fact be written as $N_{\rm gal}(E)\propto N_{\rm s}(E) \mathcal{R_{\rm SN}}\tau_{\rm gal}(E)$, which imposes $\delta+\alpha\approx 2.65$.
Since $\delta$ is inferred to be $\approx 0.3-0.6$ \cite{PAMELA14}, one finds $\alpha\approx 2.05-2.35$, slightly steeper than the DSA prediction for strong shocks.
Such a discrepancy will be discussed in \S\ref{sec:steep}, but it is remarkable how a simple homogenous diffusive model for CR transport is able to simultaneously reproduce the measured CR secondary/primary ratios, the diffuse Galactic synchrotron and $\gamma$-ray emission (see, e.g., \cite{ba12a,galprop98,galprop11,dragon08,dragon12,dragon13}), and (if $\delta\approx 0.3$) also the observed anisotropy in the arrival directions of CRs \cite{ba12b}.\\

\noindent{\bf SNR magnetic fields and the maximum CR energy.}
Another pillar of the SNR paradigm is the extent of such a universal power-law.
DSA is scale-free, and cannot predict either the minimum or the maximum energy of the spectrum of the accelerated particles. 
The minimum energy required for entering the acceleration process (\emph{injection} energy) is discussed in \S\ref{sec:inj}, while the maximum energy attainable during the SNR lifetime depends on how rapidly particles can be scattered across the shock, which in turn depends on the amplitude and the spectrum of upstream and downstream magnetic irregularities.
In SNRs magnetic fields can be factors of $10-100$ larger than in the interstellar medium (ISM), as it is inferred from the following observational facts.

{\bf i.} Young SNRs show thin non-thermal X-ray rims produced by multi-TeV electrons radiating in magnetic fields as large as a few hundred $\mu$G \cite{V+05,P+06}. 
Recent measurements in SN1006 \cite{ressler+14} and in Tycho \cite{tran+15} showed that the thickness of the rims is frequency-dependent, which allows to assess the relative importance of magnetic field damping and radiative losses and to conclude that magnetic fields are indeed amplified beyond simple compression.

{\bf ii.} X-ray hotspots in RX J1713.7--3946 show variability on a few year timescale, which may require localized magnetic fields of $\lesssim\,1$\,mG \cite{uchiyama+07}.

{\bf iii.} Fitting the SNR synchrotron spectra from radio to X-rays typically reveals electrons to be fast-cooled above the critical energy at which the loss-time equals the SNR age;  
in Tycho, for instance, this corresponds to an average downstream field of $\approx 200\,\mu$G \cite{tycho}.

{\bf iv.} The lack of detection ---within Chandra resolution--- of X-rays in front of the forward shock of SN1006 suggests that field amplification must occur in the \emph{upstream} \cite{sn1006}. 
Such an evidence challenges the scenarios in which B is amplified \emph{only} in the downstream region via turbulent dynamo processes triggered by upstream inhomogeneities (e.g., \cite{gj07,inoue+09,fraschetti13}).\\

The most intriguing aspect of such a stupendous amplification of the pre-shock magnetic field is that it is likely due to the plasma instabilities driven by the super-Alfv\'enic streaming of accelerated particles, in a non-linear chain that transfers energy from the CRs to the magnetic turbulence, and then back to the particles by enhancing their diffusion and favoring rapid energization. 
The typical ISM magnetic fluctuations correspond to a diffusion coefficient of $D_{\rm gal}(E)\simeq 10^{28}[E/(3Z\,{\rm GeV})]^{\delta}$ and would allow to achieve a very low maximum energy $\Em\lesssim 10\,$GeV at the end of the Sedov stage \cite{blasi05}; 
therefore, the ISM magnetic turbulence has way too little power at the scales resonant with CRs to allow their acceleration up to the knee.
Even if ISM magnetic fluctuations were rearranged in such a way that the mean free path for pitch-angle scattering became as small as the particle gyroradius (\emph{Bohm diffusion}, i.e., $D_{\rm B}\simeq c/3 r_L$), $\Em$ would still be limited to $\lesssim 10^5$GeV \cite{lc83a,lc83b}, a factor of $\lesssim 50$ below the observed knee.
If Bohm diffusion were achieved in the amplified magnetic fields, DSA would allow to reach energies as large as $\Ekn$ in young SNRs.

\subsection{What is missing in the standard SNR paradigm?}  
As presented in the previous section, the SNR paradigm seems to check most, if not all, of the requirements to be the ultimate theory for Galactic CR acceleration. 
In reality, it encompasses several theoretical assumptions, which have accompanied the model since its very original formulation, that have never been corroborated by first-principles calculations and/or by unequivocal observational signatures.
Some of the questions crucial to the problem are:
\begin{itemize}
\item Can DSA be as efficient as 10--20\%? What regulates such an efficiency?
\item What determines the fraction of ions and electrons that is injected into DSA?
\item How are B fields amplified in SNRs? What controls the saturation of CR-driven instabilities? 
\item How do CRs diffuse in self-generated fields, both in SNRs and in the Galaxy? 
\item Is there any observational evidence of DSA in SNRs? 
\item What determines the CR transport in the Galaxy?
\end{itemize}

\section{Collisionless Shocks: Kinetic Simulations}  \label{sec:sims}
\begin{figure}
\centering
\vspace{-0.7cm}
\includegraphics[trim=0px 5px 0px 0px, clip=true, width=0.8\textwidth]{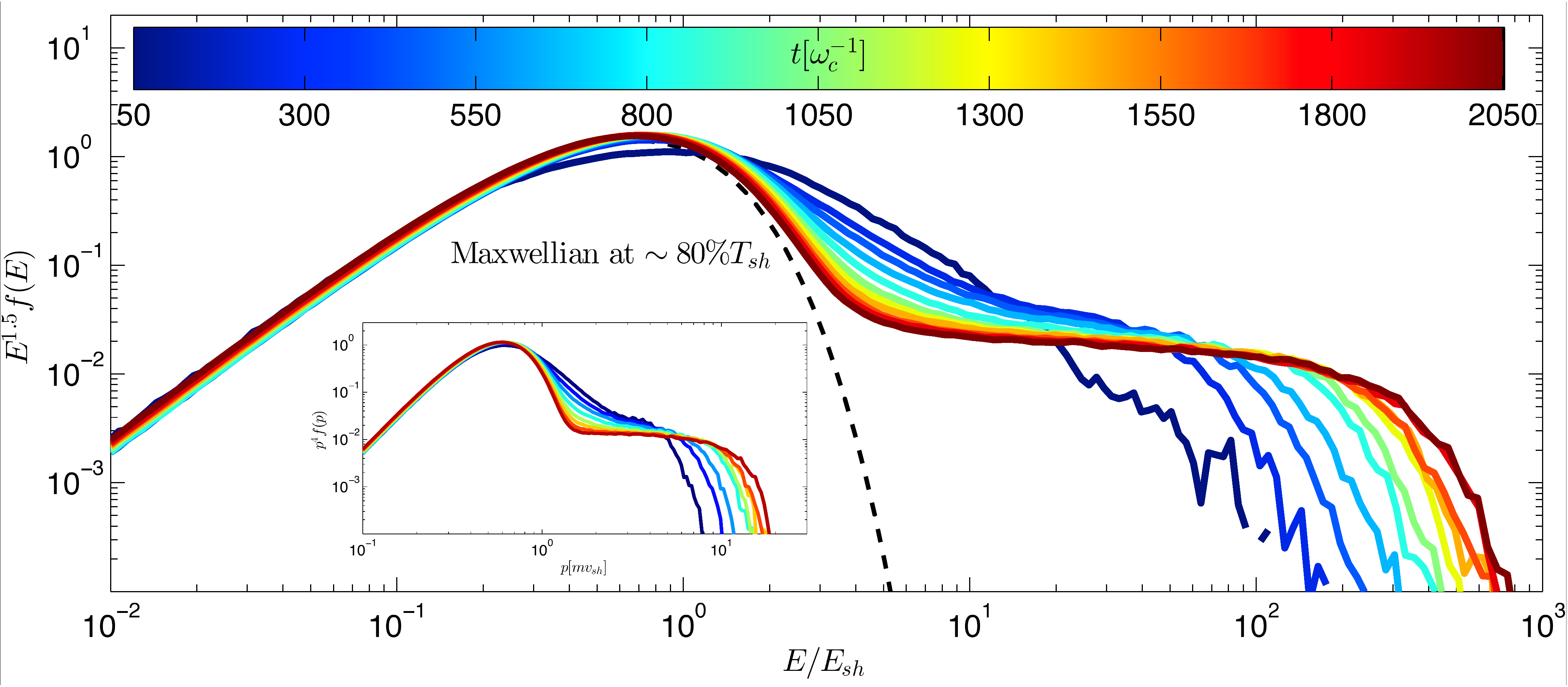}
\caption{\label{fig:p4fp}\footnotesize
Time evolution of the post-shock ion energy spectrum for a parallel shock with $M=20$, showing a non-thermal tail that stems out of the thermal distribution for $E\gtrsim 2\esh$. 
In the inset, the momentum spectrum is multiplied by $p^4$ to emphasize the agreement with DSA prediction at strong shocks.
The downstream temperature is reduced by $\sim 20\%$ with respect to standard jump conditions, because of the energy going into accelerated particles \cite{DSA}.}
\vspace{-0.3cm}
\end{figure}

To address most of these questions it is necessary to model the non-linear interplay between energetic particles and the electromagnetic fields, which is very hard to tackle analytically. 
Astrophysical plasmas are typically collisionless, i.e., their dynamics is mediated by collective interactions rather than by binary collisions, and can be fruitfully modeled \emph{ab initio} by iteratively moving particles on a grid according to the Lorentz force and self-consistently adjusting the electromagnetic fields.  
In order to mitigate the high computational cost of such particle-in-cells (PIC) simulations, one may revert to the \emph{hybrid} approach, in which electrons are considered as a massless neutralizing fluid \cite{lipatov02}, and still model shock formation, ion acceleration, and plasma instabilities self-consistently.
Hybrid simulations have been extensively used for heliospheric shocks\footnote{To give an idea, time and length scales accessible to hybrid simulations on modern supercomputers are comparable with the physical scales of the Earth's bow shock \cite{karimabadi+14}.} (e.g., \cite{giacalone+97,bs13}), but their application to astrophysical shocks has been quite limited.
SNR shocks are characterized by large sonic and Alfv\'enic ($M_A\equiv \vsh/v_A$, with $v_A=B_0/\sqrt{4\pi m n}$ the Alfv\'en velocity) Mach numbers, which makes it computationally challenging to  
capture the diffusion length of accelerated ions $D/\vsh\approx v/\vsh r_L \gg  M_A c/\omega_p$ while resolving the ion skin depth $c/\omega_p$ ($\omega_p=\sqrt{4\pi n e^2/m}$ is the ion plasma frequency and $n, e,$ and $m$ the ion density, charge, and mass).

Yet, very recently, a comprehensive analysis of ion acceleration has been performed via large 2D/3D hybrid simulations of strong shocks \cite{DSA,MFA,diffusion}.
Particularly promising is also the coupling of the hybrid technique with a MHD description of the background plasma \cite{uHybrid}.
The progress in modeling non-relativistic shocks via first-principles simulations is finally attested by the first PIC simulations showing simultaneous acceleration of both ions and electrons \cite{electrons,kato15}.


\subsection{\label{sec:hybrid} Hybrid simulations: Ion acceleration}

Large 2D and 3D hybrid simulations have been performed with the Newtonian code \emph{dHybrid} \cite{gargate+07} where the shock is set up as outlined in \cite{DSA}.
Lengths are measured in units of $c/\omega_p$, velocities normalized to the Alfv\'en speed $v_A$, and energies to $\esh\equiv m v_{\rm s}^2/2$, where $v_{\rm s}$ is the velocity of the upstream fluid in the downstream frame.
The shock strength is expressed by the Alfv\'enic Mach number $M_A$, assumed to be comparable with $M_{s}$ (both are indicated by $M$ if not otherwise specified).
The shock inclination is defined by the angle $\vartheta$ between the shock normal and the background magnetic field ${\bf B}_0$;
therefore, $\vartheta=0\deg (90\deg)$ for a parallel (perpendicular) shock.

\subsubsection{Ab-initio DSA\label{sec:DSA}}
\begin{figure}
\vspace{-0.7cm}
\centering
\includegraphics[trim=100px 0px 100px 0px, clip=true, width=.43\textwidth]{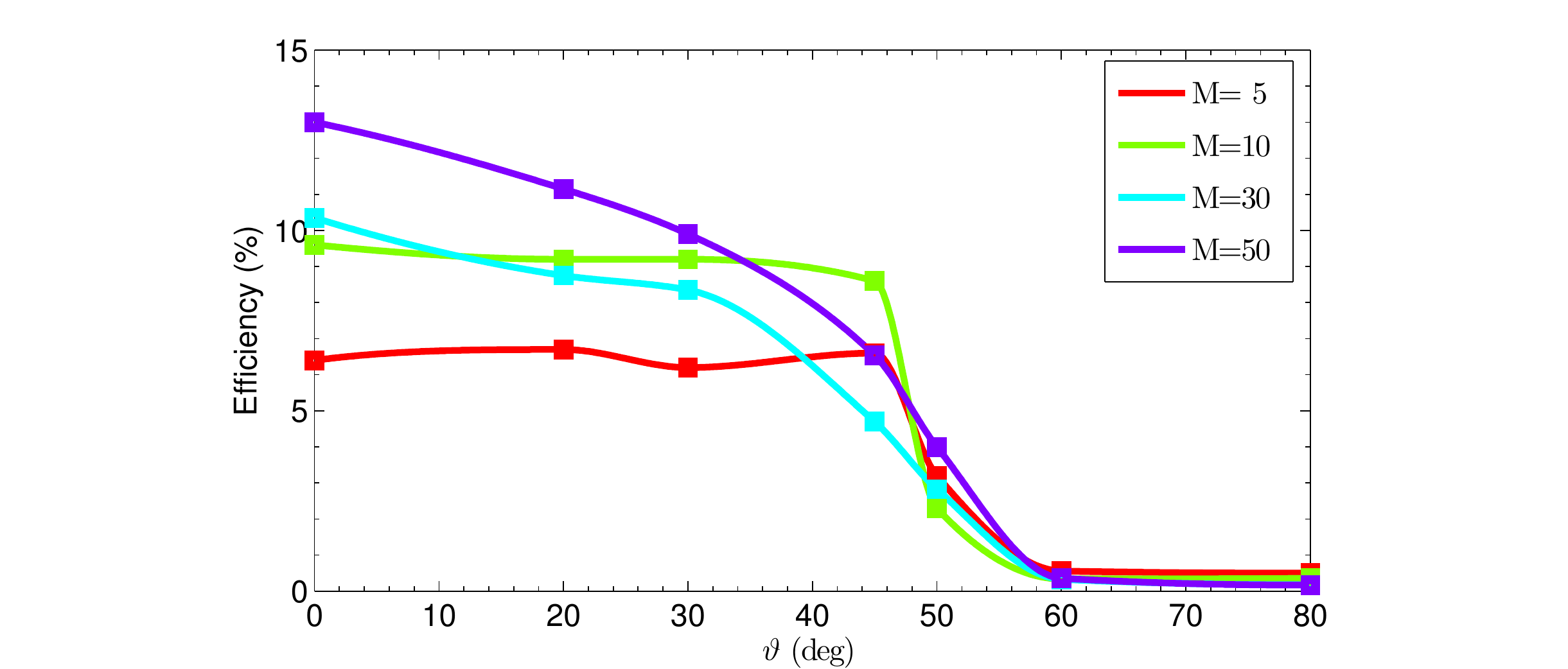}
\includegraphics[trim=110px 10px 90px 1350px, clip=true, width=.55\textwidth]{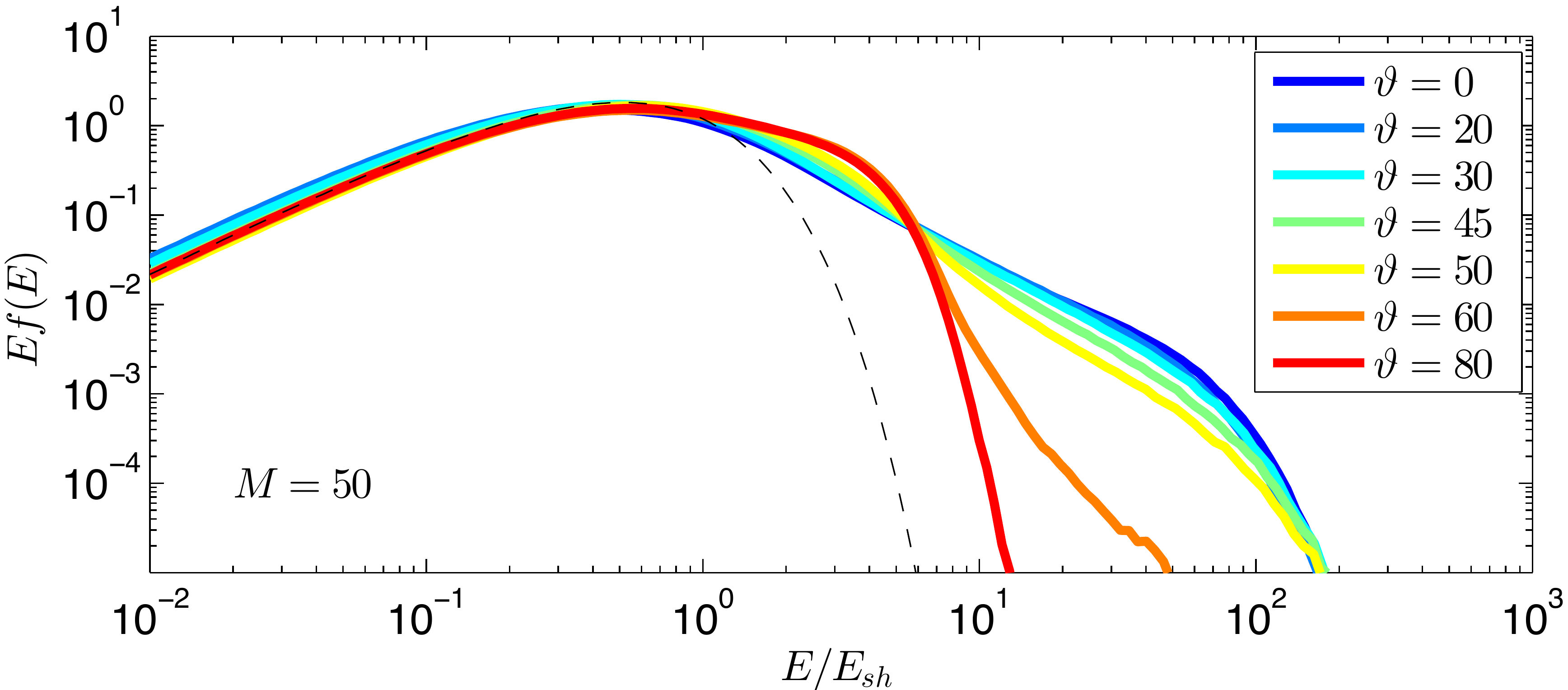}
\caption{\label{fig:eff}
\emph{Left Panel:} Fraction $\xcr$ of the downstream energy density in non-thermal particles as a function of shock inclinations and Mach numbers, $M$.
The largest acceleration efficiency ($\xcr\gtrsim 10\%$) is achieved for strong, parallel shocks, and drops for $\thbn\gtrsim 45\deg$, regardless of $M$.
\emph{Right Panel:} Post-shock particle spectra for $M=50$ and different shock obliquities, as in the legend.
The black dashed line represents the downstream Maxwellian.
Note how the non-thermal power-law tail develops only at low-inclination shocks  \cite{DSA}.
}
\vspace{-0.5cm}
\end{figure}
The kinetic simulations presented in refs.~\cite{DSA,MFA,diffusion} have been able, for the first time, to demonstrate that DSA acceleration at non-relativistic strong shocks can indeed be efficient.  
Figure \ref{fig:p4fp} shows the ion spectrum in the downstream of a parallel shock with $M=20$ \cite{DSA}; 
such a spectrum develops a non-thermal tail whose extent (corresponding to the maximum energy achieved by accelerated ions) increases with time and whose slope agrees perfectly with the DSA prediction (Eq.~\ref{eq:universal}).
In this case, a fraction of $\xcr\approx 15\%$ of the shock kinetic energy is converted into energetic ions, and the post-shock temperature is accordingly reduced with respect to Rankine--Hugoniot jump conditions.
Such a modification is an exquisite manifestation of the back-reaction of efficient CR acceleration and is usually accounted for in models of non-linear DSA (NLDSA, see \cite{je91,md01} for reviews, and \cite{comparison} for a comparison of different approaches to the problem).

The left panel of Figure \ref{fig:eff} shows $\xcr$ as a function of shock strength and inclination. 
The acceleration efficiency can be as high as $\gtrsim 15\%$ at strong, quasi-parallel shocks, and drops for $\thbn\gtrsim 45\deg$, independently of the shock Mach number. 
The right panel of Figure \ref{fig:eff}, instead, shows the ion spectra for shocks with $M=50$ and different inclinations;
the DSA non-thermal tail vanishes for quasi-perpendicular shocks, where ions gain a factor of few in energy, at most.
Also 3D simulations show the same dependence of the acceleration efficiency on $\thbn$ \cite{DSA}.

\subsubsection{\label{sec:MFA} Magnetic Field Amplification}
Since the initial formulation of the DSA theory, particle acceleration has been predicted to be associated with plasma instabilities \cite{bell78a}, in particular with the generation of magnetic turbulence at scales comparable to the gyroradii of the accelerated particles (\emph{resonant streaming instability}, see \cite{skilling75a,bell78a}). 
More recently, Bell pointed out that non-resonant, short-wavelength modes may grow faster than resonant ones (\emph{non-resonant hybrid, NRH, instability} \cite{bell04}). 
On top of these instabilities, which excite modes parallel to the background magnetic field, some transverse and  filamentary modes are expected to grow, too (e.g., \cite{rb12,filam,bykov+13}). 

\begin{figure}
\begin{center}
\vspace{-0.7cm}
\includegraphics[trim=45px 75px 60px 55px, clip=true, width=.485\textwidth]{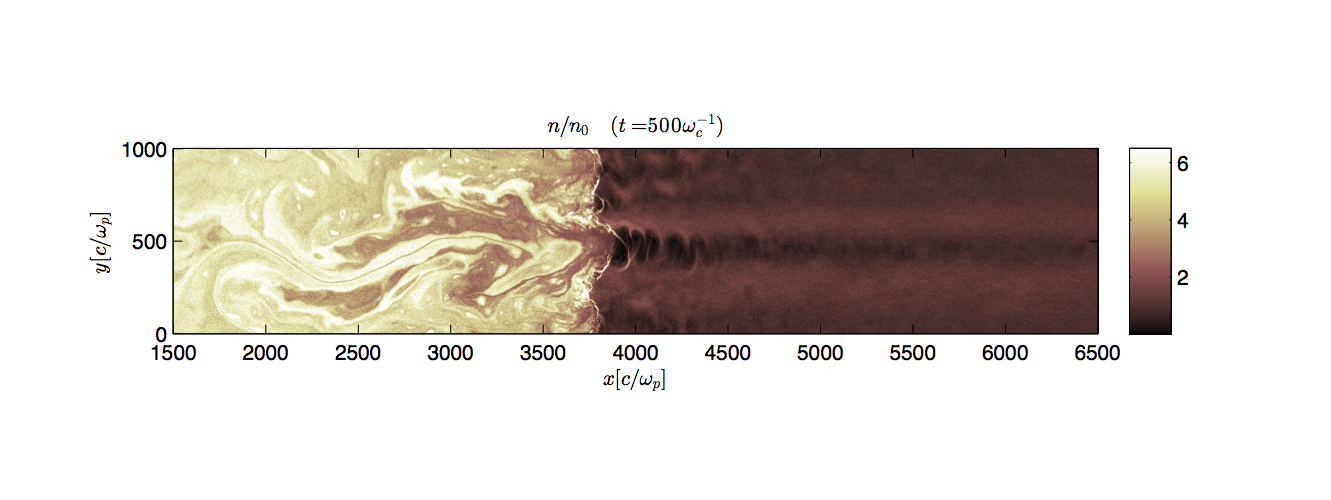}
\includegraphics[trim=35px 75px 60px 55px, clip=true, width=.485\textwidth]{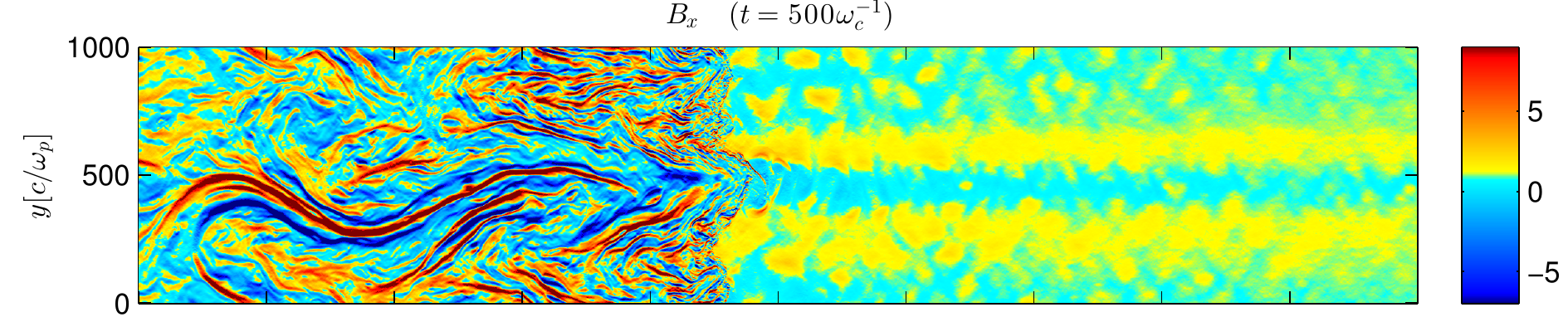}
\includegraphics[trim=45px 75px 60px 55px, clip=true, width=.485\textwidth]{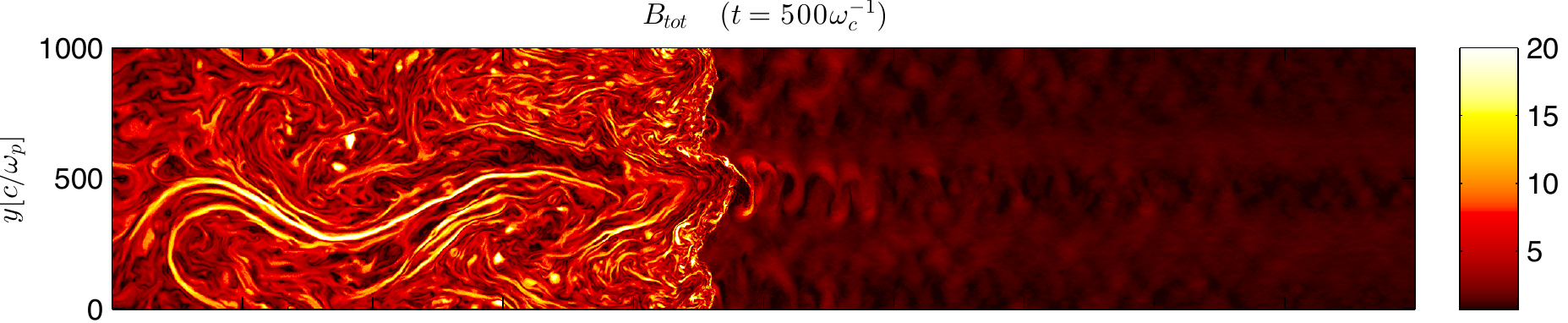}
\includegraphics[trim=35px 75px 60px 55px, clip=true, width=.485\textwidth]{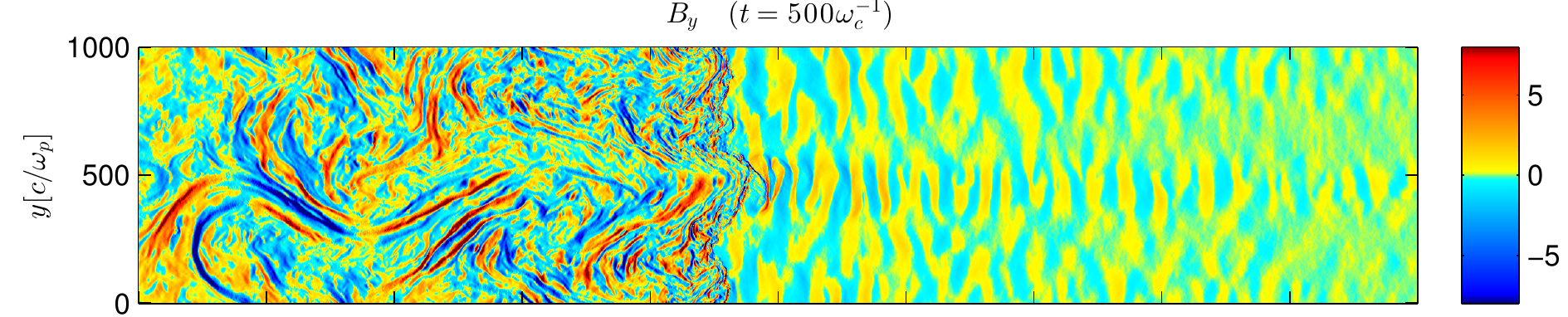}
\includegraphics[trim=35px 45px 50px 45px, clip=true, width=.485\textwidth]{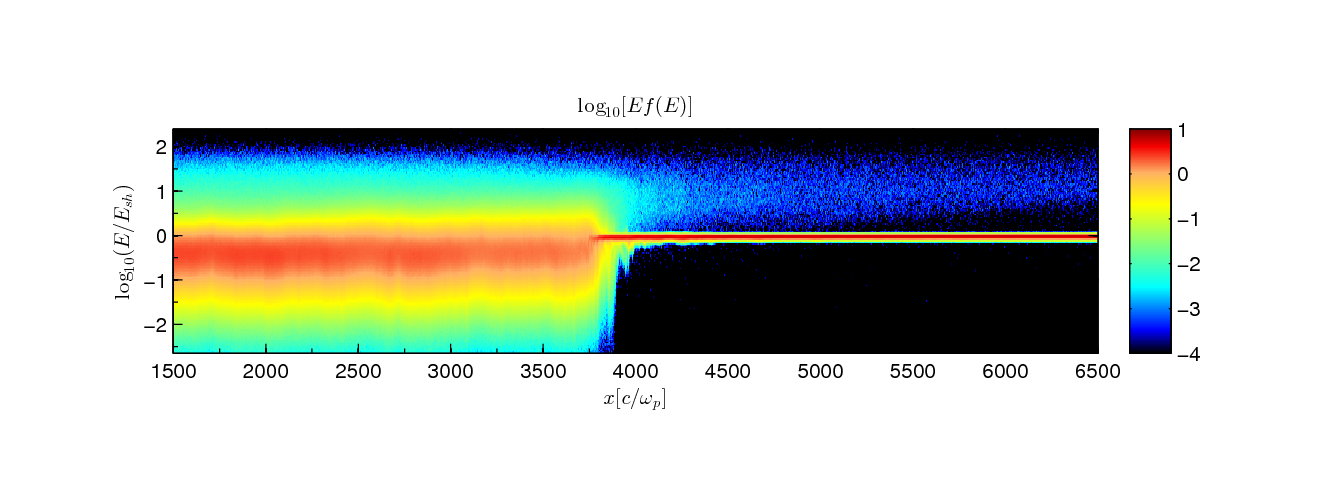} 
\includegraphics[trim=35px 45px 60px 55px, clip=true, width=.485\textwidth]{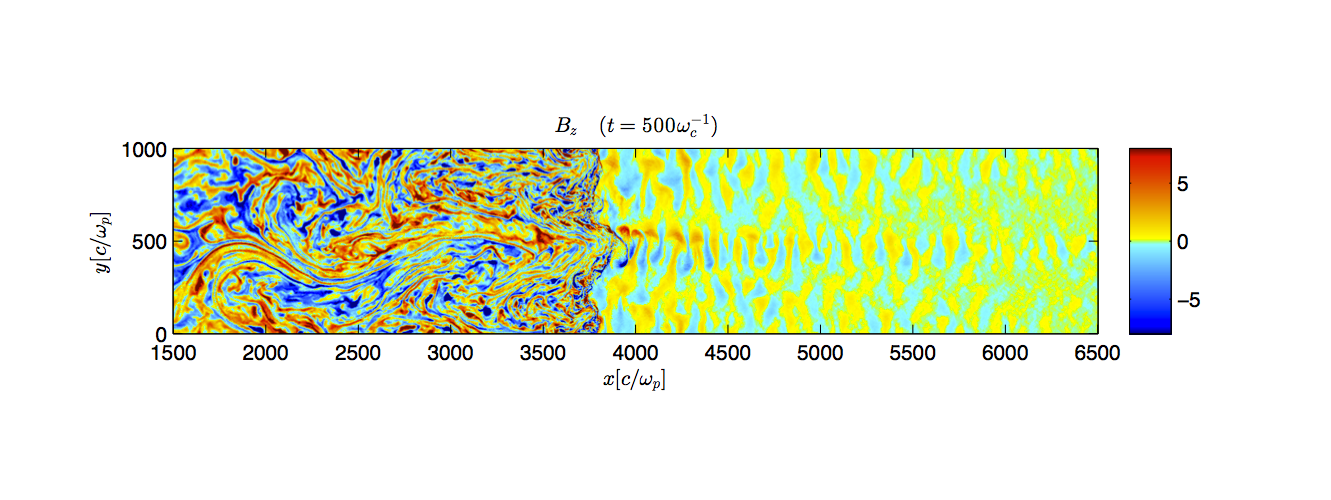}
\end{center}
\vspace{-0.5cm}
\caption{\label{fig:hybrid}\footnotesize
Output of a 2D hybrid simulation of a parallel shock with $M=30$. 
\emph{Left panels}: density; total magnetic field strength; ion $x-p_x$ phase space. \emph{Right panels}: components of the magnetic field ($B_x$, $B_y$, $B_z$). 
Such a rich structure is entirely generated by instabilities driven by accelerated ions diffusing in the upstream (to the right in the figures) \cite{filam}.}
\vspace{-0.5cm}
\end{figure}

Figure \ref{fig:hybrid} shows the structure of a parallel shock with $M=30$, with the upstream (downstream) to the right (left).
In the shock precursor, a cloud of non-thermal particles drives a current able to amplify the initial magnetic field by a factor of a few, also leading to the formation of underdense cavities filled with energetic particles and surrounded by dense filaments with strong magnetic fields  (see also \cite{rb12}).
The typical size of the cavities is comparable with the gyroradius of the highest-energy particles (a few hundred ion skin depths for the simulation in Figure \ref{fig:hybrid}).

The propagation of the shock through such an inhomogeneous medium leads to the formation of turbulent structures (via the Richtmyer--Meshkov instability), in which magnetic fields are stirred, stretched, and further amplified.
In this case, amplification via turbulent dynamo (e.g., \cite{gj07,inoue+09,inoue+13}) may become important even in the absence of large pre-existing density fluctuations.

Magnetic field generation depends on the presence of diffuse ions, hence it is more prominent at quasi-parallel shocks.
Simulations show that the maximum amplification achieved in the foreshock scales as $\delta B/B_0\propto\sqrt{M}$ and ranges from factors of a few for $M\lesssim 5$ to factors of $\gtrsim 7$ for $M\gtrsim 50$ (\cite{MFA}, fig.\ 5).
For $M\gtrsim 20$ the NRH instability grows significantly faster than the resonant one \cite{rb12,rs09,gargate+10}, exciting distinctive right-handed modes with wavelength much smaller than the gyroradius $r^*_L$ of the CRs driving the current.
Then, in the non-linear stage, an inverse cascade in $k-$space progressively channels magnetic energy into modes with increasingly small wavenumber $k$. 
The NRH instability eventually saturates when the maximally-growing mode is $k_{\max} \approx 1/r^*_L$, which effectively scatters the current ions.
This is the very reason why the resonant instability saturates already when $\delta B/B_0\sim1$ \cite{mv82} while the non-resonant one can grow up to non-linear levels before the driving current is disrupted.
For $M\lesssim 10$, $\delta B/B_0\lesssim1$ and both wave polarizations are observed, consistently with the prediction of quasi-linear theory \cite{ab09}.
The reader can refer to \cite{MFA} for a more detailed discussion of the wave spectra and the saturation of the two instabilities.

\subsubsection{\label{sec:diff}Particle Diffusion}
CRs are scattered in pitch angle by waves with resonant wavenumbers $k(p)\sim 1/r_L(p)$; in the regime of small deflections this process can be described by a diffusion coefficient. 
The most popular choice is to assume the Bohm limit, which is obtained (in the quasi-linear limit $\delta B/B_0\lesssim 1$) for an Alfv\'enic turbulence generated via resonant streaming instability by a CR distribution $\propto p^{-4}$ \cite{bell78a}.
Bohm diffusion is often heuristically extrapolated into the regime of strong field amplification, but such a prescription used to lack a solid justification.  

\begin{figure}\centering
\vspace{-0.7cm}
\includegraphics[trim=0px 50px 0px 260px, clip=true, width=.485\textwidth]{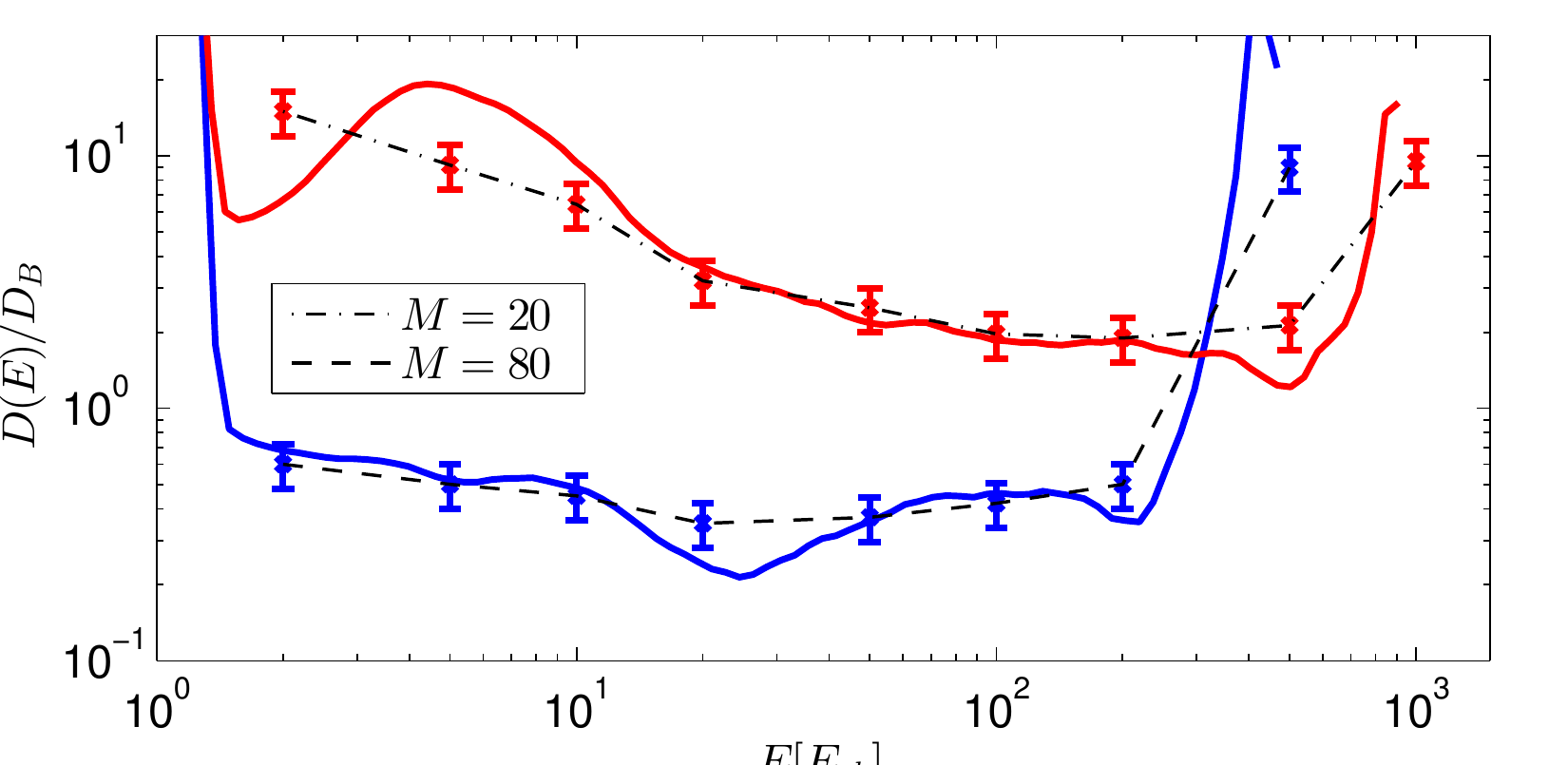}
\includegraphics[trim=30px 0px 40px 15px, clip=true, width=.485\textwidth]{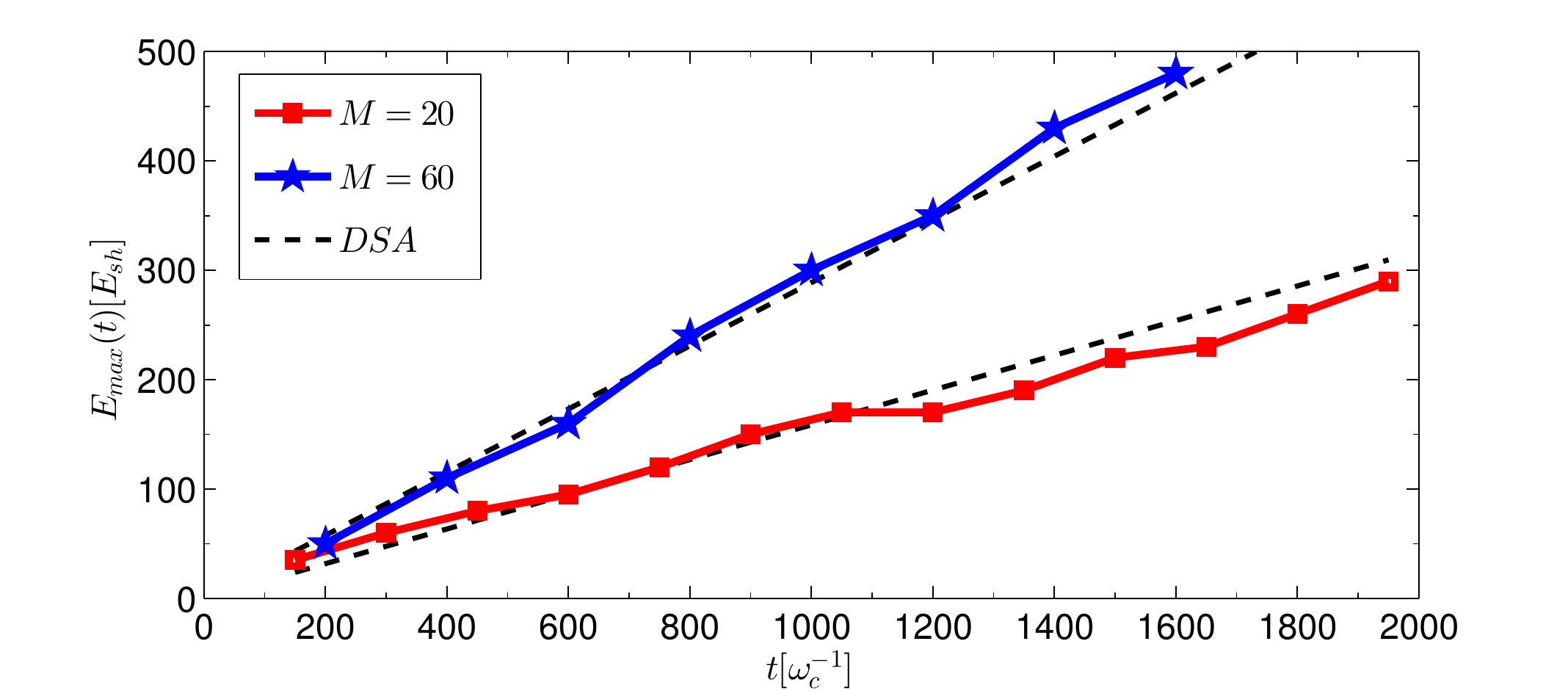}
\caption{\label{fig:cd}\footnotesize
{\emph{Left panel}: Diffusion coefficient, normalized to Bohm, immediately in front of the shock for hybrid simulations of shocks with $M=20, 80$.
\emph{Right panel}: Time evolution of the maximum ion energy for parallel shocks with $M=20,60$, compared with the DSA prediction  (dashed lines) \cite{diffusion}.}
}
\vspace{-0.5cm}
\end{figure}

Global hybrid simulations allow to reconstruct CR diffusion in different regions of the shock, either by using an analytical procedure based on the extent of the CR distribution in the upstream or by tracking individual particles \cite{diffusion}.
The two methods return consistent results, as shown in the left panel of Figure \ref{fig:cd} (see \cite{diffusion} for more details).
When magnetic field amplification occurs in the quasi-linear regime ($M\lesssim20$), particle scattering is well described by the diffusion coefficient self-generated via resonant instability \cite{bell78a,ab06}, where the scattering rate depends on the magnetic power in resonant waves.
For stronger shocks, instead, $D(E)$ is roughly proportional to the Bohm coefficient and its \emph{overall} normalization depends on the level of magnetic field amplification $\delta B/B_0\gtrsim 1$ (see also \cite{rb13}).
Such a scaling is determined by the fact that far upstream the spectrum of the excited magnetic turbulence (\cite{MFA}, figs.\ 6 and 7) peaks at relatively large wavelengths, comparable with the gyroradius of the highest-energy ions.  

The effective scattering rate is also imprinted in the time evolution of $\Em$. 
The right panel of Figure \ref{fig:cd} shows such an evolution, which is linear with time with a slope inversely proportional to the measured diffusion coefficient (dashed lines), as expected for DSA (e.g., \cite{drury83,lc83a,maximum}).

\subsubsection{\label{sec:inj} A theory of ion injection}

Explaining the correlation between ion acceleration and shock obliquity requires understanding the conditions necessary for thermal particles to be injected into DSA.
High-resolution hybrid simulations  show that all the ions that eventually achieve large energies are reflected by the shock potential barrier at their first shock encounter \cite{injection}. 
The sharp shock transition (few ion skin depths) is associated with compression and pressure increase (overshoot), and with a cross-shock electric potential that generates an upstream-directed electric field.
At quasi-parallel shocks, the coherent reflection of impinging ions induces a shock reformation about one gyroradius upstream of the shock (e.g., \cite{Lee+04, su+12, injection});
because of such a reformation, the height of the barrier fluctuates on a Larmor timescale ($\sim \pi/\omega_c$), which sets the period for ion injection.

At any shock reformation cycle, about $25\%$ of the incoming ions are reflected, but not all of them enter DSA.  
Ions impinging on the shock may turn into:  
i) \emph{thermal ions}, which encounter a ``low'' barrier too weak to reflect them and immediately cross downstream;
ii) \emph{supra-thermal ions}, which are initially reflected by a ``high'' barrier, but are advected downstream during their first few gyrations around the shock, achieving a maximum energy $E\lesssim \Ei \approx10\, E_{\rm sh}$ via shock drift acceleration (SDA, e.g., \cite{scholer90,su+12});
iii) \emph{non-thermal ions}, which are reflected, energized via several cycles of SDA, and eventually achieve an energy $E\gtrsim \Ei$ that allows them to escape upstream. 
Only non-thermal ions are really injected into the DSA process, since they must rely on diffusion ---possibly on self-generated turbulence--- to get back to the shock.
The existence of supra-thermal ions at quasi-perpendicular shocks, which in simulations do not show DSA tails (Figure \ref{fig:eff}), demonstrates that reflection is necessary but not sufficient condition for DSA injection.

\begin{figure}\centering
\vspace{-0.7cm}
\includegraphics[trim=68px 260px 62px 260px, clip=true, width=.47\textwidth]{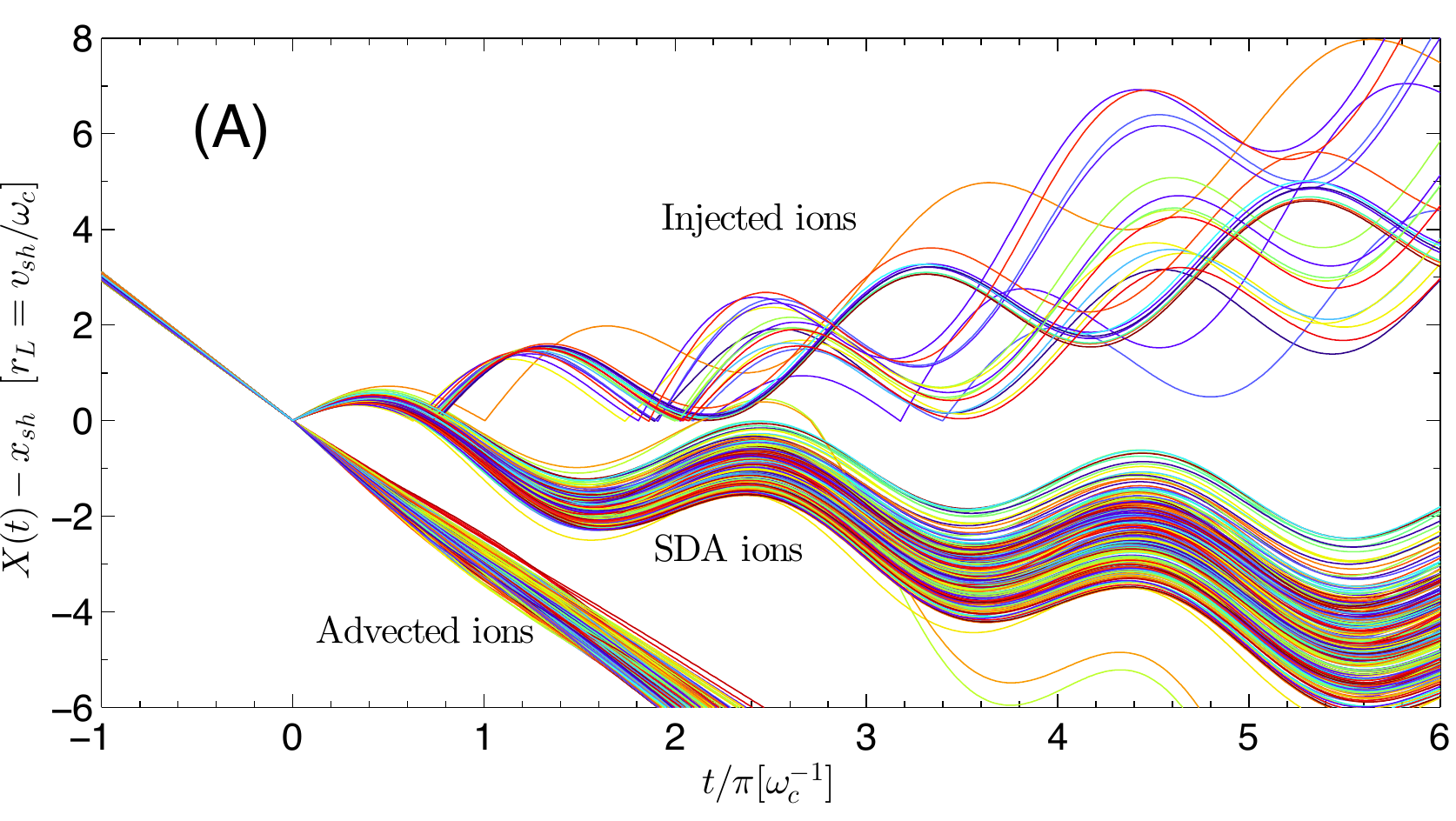}
\includegraphics[trim=12px 250px 10px 240px, clip=true, width=.5\textwidth]{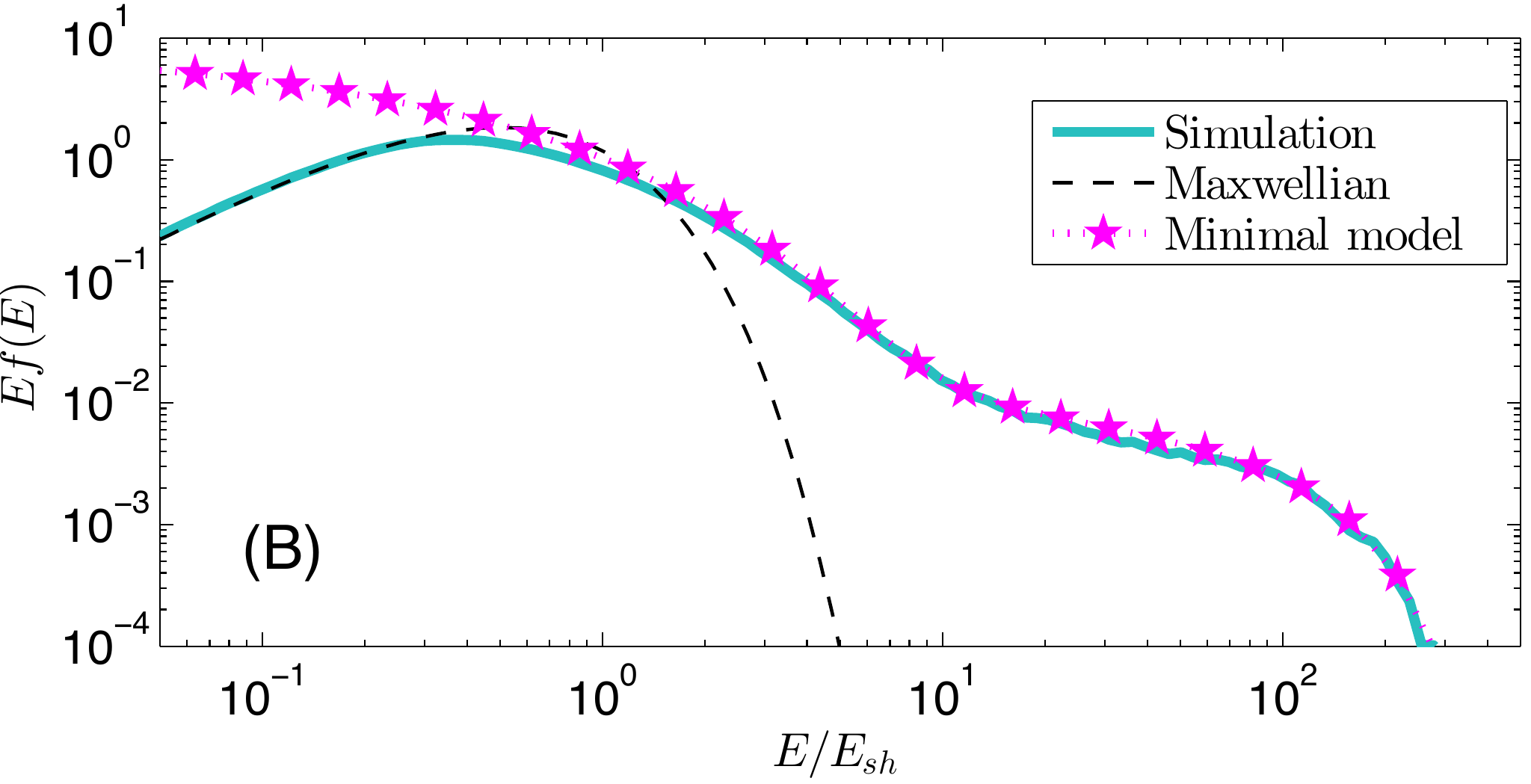}
\caption{\label{fig:traj} 
\emph{Left Panel:} Trajectories of test-particles impinging at random times on a periodically-reforming shock with $M=10$ and $\thbn\simeq 45\deg \pm 2\deg$ \cite{injection}; for each ion, $t=0$ corresponds to the first shock encounter.
Ions may either not reflect (``Advected ions''), or experience SDA before ending up downstream (``SDA ions''), or escape upstream after a few reflections (``Injected ions'').
\emph{Right Panel:} Post-shock ion spectrum for a parallel shock with $M=10$, as obtained in simulations and compared with the minimal model outlined in ref.~\cite{injection}.}
\vspace{-0.5cm}
\end{figure}

By generalizing the formalism of ref.~\cite{BS84}, it is also possible to calculate the minimum energy $\Ei$ needed to escape upstream of a shock with a given inclination.
The main results are (see \S3 of \cite{injection} for more details):
i) cold ions can be directly injected only if $\thbn\lesssim 30\deg$; 
ii) ion injection for $\thbn\gtrsim 30\deg$ requires SDA pre-energization.
The number of cycles needed to achieve $\Ei(\thbn)$ increases with $\thbn$ and, since gyrating ions have a finite probability to encounter the barrier in the low state and be advected downstream, the fraction of ions that can perform an increasingly large number of cycles is exponentially suppressed;
iii) for inclinations $30\deg\lesssim \thbn\lesssim 55\deg$, $\mathcal{N}\approx 2-3$ SDA cycles are needed to reach $\Ei\approx 10\,\esh$ and $\sim 0.25^\mathcal{N}\sim1\%$ of the incoming particles is injected. Above $\thbn\sim 60\deg$, $\mathcal{N}\gtrsim 4$, and the fraction of ions that escape upstream goes to zero quite rapidly.
DSA-efficient shocks always converge to a configuration where an effective inclination $\thbn\approx 45\deg$ is achieved because of the non-linear field amplification in the precursor \cite{MFA}.
These findings can be encapsulated in a \emph{minimal model for ion injection} \cite{injection}, which accounts for shock reformation and reflection at the shock barrier, and reproduces the fraction of ions in the supra-thermal and non-thermal distributions, as well as their phase-space distribution (see Figure \ref{fig:traj}).

\subsection{PIC simulations: Electron acceleration \label{sec:PIC}}
The injection of electrons into DSA has traditionally been an outstanding problem since they have smaller gyroradii compared to ions and the electrostatic barrier crucial for ion reflection is instead pernicious to electrons.
Despite several different mechanisms have been proposed to address these issues (e.g., \cite{ah07,rs11,matsumoto+15}), a comprehensive theory of electron injection is still missing. 

\begin{figure}
\centering
\vspace{-8mm}
\includegraphics[trim=0px 0px 0px 0px, clip=true, width=0.44\textwidth]{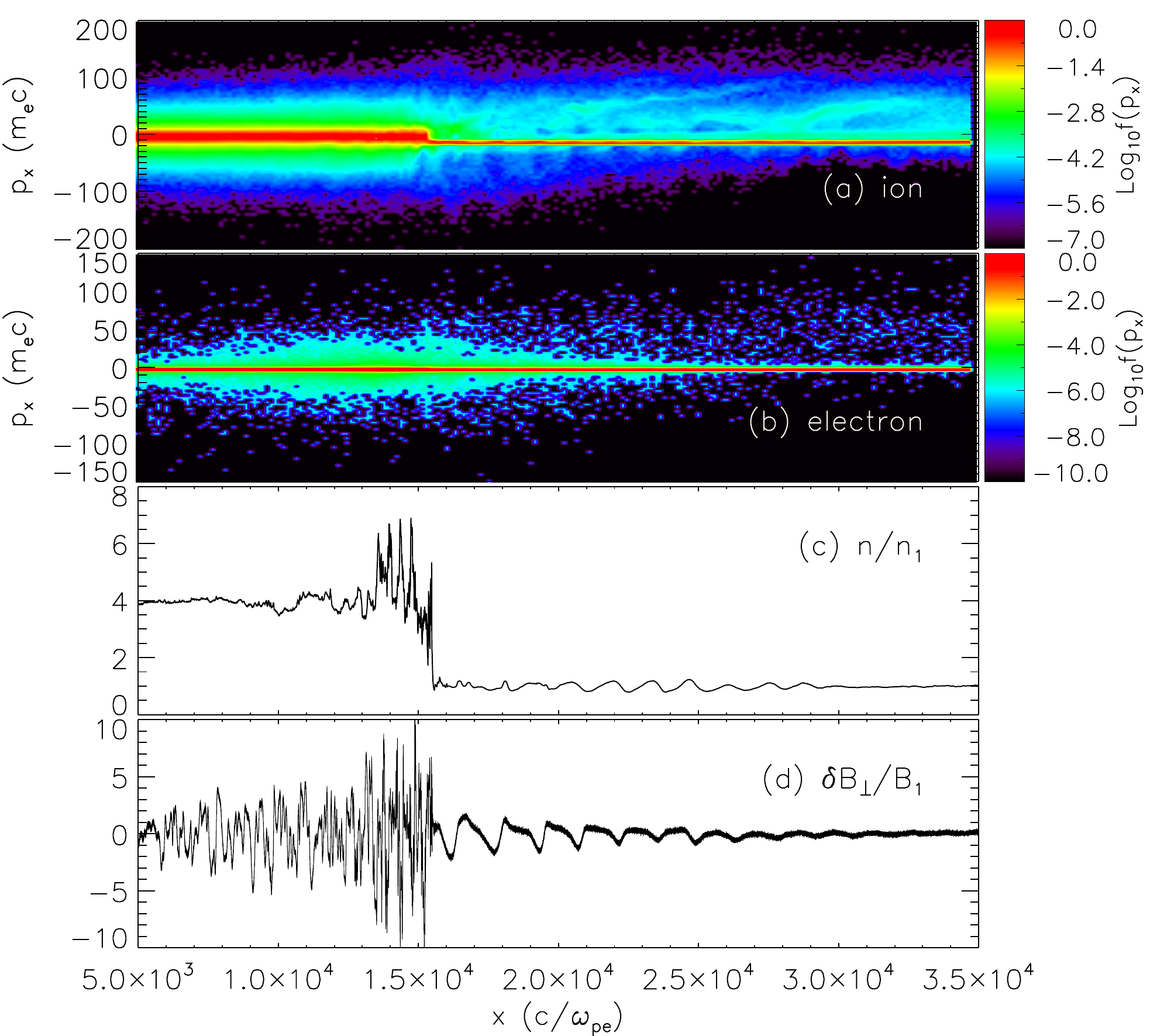}
\includegraphics[trim=0px 0px 20px 0px, clip=true,width=0.55\textwidth]{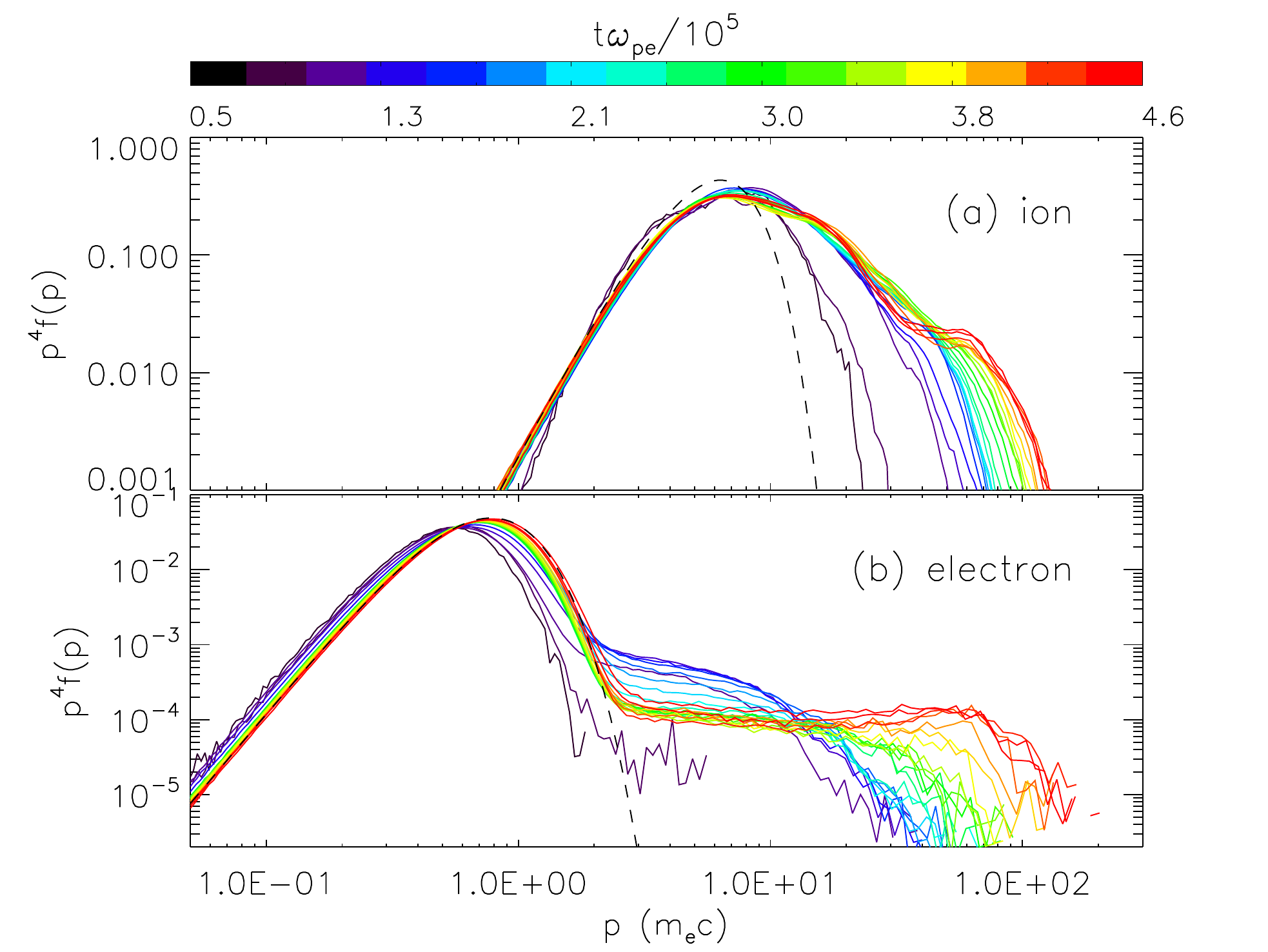}
\caption{\label{fig:ele} \footnotesize
\emph{Left Panel:} Proton (a) and electron (b) $x-p_x$ phase space distributions, density profile (c), and transverse component of ${\bf B}$ (d) for a shock with $v_{\rm sh}=0.1c$, $\thbn=30^\circ$, and $m_p/m_e=100$.
Energetic protons and electrons diffuse ahead of the shock, amplifying the upstream magnetic field.
\emph{Right Panel:} Evolution of the downstream momentum distributions for (a) protons and (b) electrons.
The dashed lines correspond to thermal distributions \cite{electrons}.
}
\vspace{-4mm}
\end{figure}

Only the computationally-expensive PIC approach can study electron acceleration ab initio. 
This year the first (1D) simulations of non-relativistic shocks that show simultaneous acceleration of both electrons and ions appeared in the literature \cite{electrons,kato15};
in particular, ref.~\cite{electrons} shows that at quasi-parallel shocks both species develop the DSA power-law tail $\propto p^{-4}$, allowing the first self-consistent measurement of the electron/proton ratio in accelerated particles.
The reported value of  $\Kep\approx 10^{-3}$ is not too different from those inferred from multi-wavelength observations of young SNRs and in Galactic CRs.
Figure \ref{fig:ele} shows ion and electron phase-space and spectra, as well as density and self-generated magnetic field profiles for a shock with $\vsh=0.1c$, $M=20$, and $\thbn=30\deg$.

Electron acceleration feeds on the NRH modes excited by energetic ions, which increases the effective inclination of ${\bf B}$ at the shock and induce electron reflection because of magnetic mirroring\footnote{Ions do not conserve their magnetic moment since the shock is reforming on their Larmor time scale (\S\ref{sec:inj}).}. 
Non-resonant modes have the right polarization to effectively scatter electrons, too.
Preliminary results [\emph{Caprioli et al., in prog.}] show that the electrons energized in the shock foot (e.g., \cite{lembege02a, ms06, electrons}) can achieve velocities large enough to be injected even for $\thbn\gtrsim 45\deg$, despite ions can not.
Electron reflection and a hint of a Fermi-like process at oblique shocks has been reported also for the weak shocks ($M_A\gg M_s\gtrsim 1$) typical of  galaxy clusters \cite{guo+15a,guo+15b}.

\subsection{Comparisons with observations}
\begin{figure}\centering
\vspace{-.7cm}
\includegraphics[trim=0px 5px 0px 90px, clip=true, width=1\textwidth]{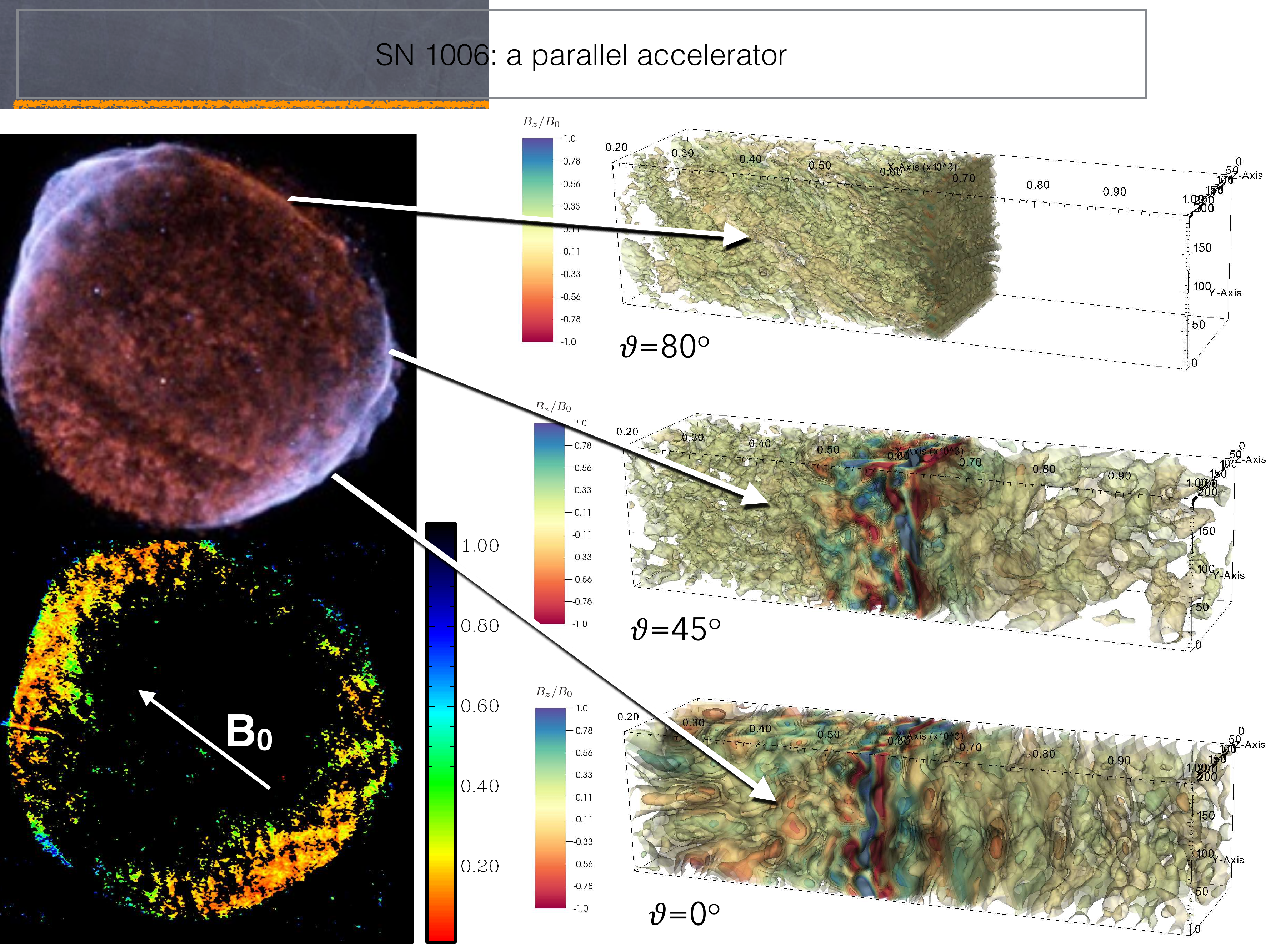}
\caption{\label{fig:sn1006}\footnotesize
{The top left panel shows the thermal (red) and non-thermal (light blue)  X-ray emission from SN1006, while the bottom left panel shows the corresponding degree of polarization in the radio band, as in the legend \cite{reynoso+13};   
the arrow shows the fiducial direction of the background Galactic magnetic field ${\bf B}_0$. 
Right panels show the self-generated magnetic fields for shocks of different inclinations, obtained in the hybrid simulations in ref.~\cite{MFA}.
The quasi-parallel regions show little polarization and enhanced synchrotron emissivity, consistent with simulation outputs.}
}\vspace{-.5cm}
\end{figure}

A natural question is whether such findings have any observational counterpart in SNRs or in other non-relativistic shocks with known geometry. 
The typical coherence length of the Galactic magnetic field is of order of 10--50\,pc, comparable with the radius of most SNRs, with two notable exceptions: SN1006, which lies significantly above the Galactic plane, where the field coherence length should be the larger than in the disk, and G1.9+0.3, the youngest (and likely smallest) SNR in the Milky Way \cite{reynolds+08}. 
These SNRs show a bilateral symmetry,  found also in some older SNRs, that may correlate with the geometry of the background magnetic field (e.g., \cite{rothenflug+04}).
Recent observations of SN1006 (Figure \ref{fig:sn1006}) show that the radio emission from the synchrotron-bright polar caps has a low degree of polarization, implying the presence of a strong and turbulent magnetic field; 
such regions are also inferred to be those parallel to the large scale ${\bf B}_0$\footnote{Starlight polarization in the direction of SN1006 is consistent with this picture, too [B. Draine, priv.\ comm.]} \cite{reynoso+13}, in qualitative agreement with the trend of magnetic field amplification seen in simulations \cite{DSA,MFA}.
Finally, the Tycho's ``stripes'' \cite{eriksen+11} and the bright knots observed in RX J1713.7--3946 may be signatures of the filamentation instability operating in the localized regions where the shock is parallel \cite{filam}.

On a much larger scale, the weak shocks in galaxy cluster can also be used to probe the conditions conducive to electron and ion acceleration (e.g., \cite{zandanel+15,vazza+15} and the review \cite{bj14}). 
The prominent radio emission of giant halos requires in some cases large electron acceleration efficiencies of $\xi_{\rm e}\simeq \Kep \xcr\gtrsim 1\%$; 
also the non-detection of $\gamma$--rays with Fermi-LAT favors values of $\Kep$ significantly larger than those inferred in SNRs.
The high degree of polarization of very extended ($\sim$Mpc) shocks (e.g., \cite{vanweereen+12,ogrean+13}) suggests either a mostly-perpendicular shock geometry or that the radio emission is dominated by the oblique regions of the shock, where ion injection is disfavored.

Finally, at the terrestrial bow shock, accelerated (diffuse) ions are only detected in the quasi-parallel regions, while energetic electrons are observed upstream also for more oblique geometries (e.g., \cite{wilsoniii15b} and refs.\ therein).
Such a transition from an ion- to electron-foreshock modulated by the shock inclination is consistent with kinetic simulations without pre-existing turbulence (\S\ref{sec:MFA}).


\section{Gamma-ray observations of SNRs\label{sec:gamma}}
$\gamma$-ray observations of SNRs have been performed both via imaging atmospheric Cherenkov telescopes (IACTs, such as HESS, MAGIC, and VERITAS) in the TeV energy range and via satellites (Fermi and AGILE) in the GeV band;
also water Cherenkov observatories (such as HAWC) can probe multi-TeV $\gamma-$rays, but they have not reported any SNR detection so far. 
For recent reviews on $\gamma$-ray astronomy, including SNR observations, see refs.\ \cite{lemoine-goumard15,buehler15,aharonian15}

\subsection{\label{sec:pev} Hunt for hadronic PeVatrons}
$\gamma$-ray emission may be either leptonic (relativistic bremsstrahlung and inverse-Compton scattering, IC) or hadronic ($\pi^0$ decay).
SNR spectra typically exhibit cutoffs around $\Ecut\approx 10$ TeV, which implies the presence either of protons with $\Emp\sim 7\,\Ecut$, or of electrons with $\Eme \sim 10\, {\rm TeV}\,\sqrt{\Ecut/{\rm TeV}}$ in the case of IC of CMB photons in the Thompson regime;
if $\Eme\gtrsim 150$\,TeV, the onset of the Klein--Nishina regime enforces $\Ecut\lesssim \Eme$.
Any Galactic accelerator able to produce hadrons\footnote{Pulsar wind nebulae do show acceleration of electrons and positrons up to PeV energies.} above a few PeV (PeVatron) should also emit $\gamma$--rays above a few hundred TeV. 
At the time of writing, no PeVatron has been associated with Galactic SNRs, even if few objects show no clear evidence of a high-energy cutoff, such as HESS J1641--463 \cite{oya15}.

Nevertheless, such a lack of detection is not inconsistent with the SNR paradigm.
For decades the maximum CR energy has been thought to be achieved at the end of the ejecta-dominated stage, just before the shock slows down because of the inertia of the swept-up material (e.g., \cite{lc83a,maximum}).
Now that the crucial role of the NRH instability has been attested, the most refined models for the evolution of $\Em$ find that multi-PeV energies can only be achieved in powerful SNe ($\gtrsim 10^{51}$\,erg) exploding in dense pre-SN winds and that SNRs might act as PeVatrons only for $\lesssim 50-100$\,yr  \cite{sb13,cardillo+15}.
Historical SNRs such as Cas A, Tycho, and SN1006 are no longer expected to be PeVatrons; 
G1.9+0.3 is only $\sim120$\,yr old, but its line of sight is too close to the Galactic center for optimal $\gamma$-ray measurements.
Even with more optimistic scalings of $\Em$ with time, the number of young Galactic SNRs that fall within the sensitivity of present $\gamma$-ray telescopes is consistent with no detections \cite{cristofari+13}. 
It is indeed possible that the Milky Way already hosts (or will host) a very young SNR that may be detectable in the next few years, especially thanks to CTA. 

\subsection{\label{sec:had} Hadronic or leptonic?}
Radio observations unequivocally attest to the presence of relativistic electrons in SNRs.
The direct evidence of hadronic acceleration, instead, may be revealed by $\gamma$-rays produced via the decay of neutral pions originated in nuclear collisions between CRs and the thermal plasma (e.g., \cite{DAV94}).
For a parent particle spectrum of $E^{-\alpha}$, bremsstrahlung and pion decay return photon energy spectra of $\nu F(\nu)\propto \nu^{2-\alpha}$, while IC returns a flatter spectrum of $\nu F(\nu)\propto \nu^{(\alpha-1)/2}$.
For the fiducial DSA index of $\alpha=2$ and for $\Kep\lesssim 10^{-3}$, IC and pion decay provide a comparable flux in the TeV range, assuming standard ISM density and photon background (e.g., \cite{ellison+07}), while the GeV flux is typically dominated by pion decay;
bremsstrahlung may be important only if $\Kep\gg 10^{-3}$. 
IACTs almost invaribaly measure photons produced by particles close to the cutoff, which makes it very difficult to disentangle the nature of the emission process.
In the few cases in which both GeV and TeV data are available for a remnant, instead, it may be possible to assess the leptonic/hadronic origin of the emission by fitting the multi-wavelength SNR emission from radio to TeV $\gamma$--rays.

The overall spectral slope from GeV to TeV energies is by itself an indicator of the nature of the emission: 
 spectra $\propto E^{-2}$ or steeper should be produced by pion decay, since leptonic processes would require either a very large $\Kep$ (for bremsstrahlung) or a very steep electron spectra (for IC), inconsistent with radio observations (e.g., \cite{gamma}).
On the other hand, SNRs exhibiting hard $\gamma$-ray spectra (e.g., RX J1713.7--3946, RCW 86, Vela Jr., SN1006) are more naturally explained in the leptonic scenario (e.g., \cite{gio+09,za10,ellison+10,acero+15}). 
A natural caveat comes from the finite resolution of $\gamma$-ray instruments, which often cannot resolve shock inhomogeneities or even the presence of MCs, which may allow for a much more complicated phenomenology (e.g., \cite{berezhko+13,ga14}).

The spectra of two middle-age SNRs (W44 and IC443) show the characteristic low-energy cutoff below the $\pi^0$ mass, which can be considered the smoking gun for the hadronic nature of the emission \cite{pionbump,IC443AGILE}.
Yet, the question remains whether such CRs have been freshly accelerated, or are diffuse CRs merely re-energized by the recent shock passage, as also suggested but the spectral slope quite similar to that of diffuse CRs. 

The best evidence for \emph{local} CR acceleration in SNRs comes from the multi-wavelength emission of single SNRs.
Tycho (SN1572) is arguably the best candidate for such an analysis: 
is the remnant of a type-Ia SN, so that age, explosion energy, and ejecta mass are well constrained; 
its light echo has been measured, which returns a distance of $\sim 3$\,kpc; 
and its morphology is quite spherically symmetric and well resolved in the radio and X-rays.
The groups that have investigated its morphology and multi-wavelength emission by means of the most refined theoretical models  all concluded that it must be an hadronic accelerator, with an efficiency of $\sim 15\%$ \cite{tycho,slane+14,berezhko+13}. 

The number of Fermi detection of TeV-emitting SNRs is steadly increasing \cite{gamma,brandt+15,snrcat}; 
therefore, it is tempting to work out a population synthesis. 
Older SNRs tend to show spectra $\propto E^{-2.5}$ or steeper (e.g., IC443, W44, W28, W51, etc.) and are almost invariably associated with MCs, while younger remnants may show either rather steep (e.g., Tycho and Cas A) or flat spectra (e.g., RX J1713.7--3946, RCW 86, Vela Jr., and SN1006). 
Such a trend may either reflect the time evolution of $\Kep$ and/or $\xcr$ or just be driven by the environment: 
the same SNR, with the same content of relativistic electrons and ions, may look more 
``hadronic'' in the presence of dense gas reservoirs (e.g., warm phase ISM, MCs) or more ``leptonic'' in the presence of an intense IC background (e.g., infra-red and optical photons in star-forming regions).

It is plausible that old SNRs can be detected in the $\gamma$-rays only if there is a large reservoir of targets, while the dominant emission mechanism in young SNRs may depend on their circumstellar medium, which for core-collapse SNe is a complex mixture of underdense bubbles and dense MCs. 
The cases of Tycho and SN1006 are paradigmatic: they both have a type-Ia progenitor and are just at the beginning of the Sedov stage, but SN1006 looks more leptonic because it is expanding in a much more rarefied medium (about 0.05 vs 1 protons cm$^{-3}$).
IC would likely dominate over $\pi^0$ decay also in Tycho if the upstream density were smaller by a factor of $\sim$20 (\cite{tycho}, fig.\ 11).

\subsection{\label{sec:steep} The origin of steep spectra}
$\gamma$-ray spectra steeper than $E^{-2}$ may be good indicators of hadronic emission, but are also ad odds with the prediction of standard DSA for strong shocks, and even more inconsistent with the \emph{flattening} expected when CR acceleration is efficient \cite{je91,md01}.
The NLDSA theory includes the backreaction of CRs (and amplified magnetic fields \cite{jumpl,jumpkin}) on the shock dynamics, predicting for large  $\xcr$ concave CR spectra as flat as $E^{-1.5}$ at the highest energies.
A simple way of thinking to this effect is considering  Eq.~\ref{eq:universal} with the compression ratio $r\simeq 7$ established by the CR relativistic fluid with adiabatic index 4/3 (e.g., \cite{dv81a, dv81b,be99}). 
A few possible solutions have been put forward to solve this apparent discrepancy between standard DSA and observations.

\textbf{Magnetic feedback.} 
In principle, CRs do not feel the fluid compression ratio, but rather the compression ratio of the magnetic fluctuations they are coupled with \cite{bell78a}. 
The velocity of such fluctuations is of the order of the Alfv\'en velocity and is typically much smaller than the fluid velocity in the shock reference frame.
In the presence of magnetic field amplification, instead,  this effect may become non-negligible if $v_A\propto \delta B/B_0$ and induce an effective compression ratio $\tilde{r}\lesssim r$ for the CRs, which in turns leads to steeper spectra (Eq.~\ref{eq:universal}) \cite{zp08b,crspectrum,gamma}.
The non-linear balance between the flattening induced by efficient CR acceleration and the magnetic feedback (which also increases with $\xcr$) is discussed in \cite{efficiency}.
For typical SNR parameters, $\xcr\sim 10\%$ and the level of magnetic field amplification inferred in SNRs  generally lead to spectral indexes $\alpha\approx 2.2-2.3$ at the beginning of the Sedov stage, exactly as required for explaining Tycho's spectrum \cite{tycho,slane+14}.
Even if this mechanism seems to work for phenomenological purposes, the assumed scaling of the wave phase velocities in the very non-linear regime  ($\delta B/B_0\gtrsim 100$) relevant for young SNRs has not been convincingly proven, yet.

\textbf{Neutral feedback.}
Most of the $\gamma$-ray--bright SNRs are either associated with dense MCs (e.g., \cite{cs10}) or propagate into partially-ionized media, as revealed by their prominent H$\alpha$ emission (e.g., \cite{cr78}). 
The high resolution of optical telescopes often allows to detect both a broad and a narrow Balmer line, whose widths are determined by the downstream and upstream plasma temperature when ions and neutrals are coupled via charge exchange.
H$\alpha$ emission is the only way of directly probing ion temperature in SNRs and measurements have revealed interesting deviations with respect to the standard predictions of gaseous shocks: sometimes narrow lines are too broad to be consistent with quasi-neutral gas in the first place (e.g., \cite{vanadelsberg+08,heng10}) and often  broad lines are anomalously-narrow, which initially suggested that a large fraction of shock energy ended up in CRs rather than in heat \cite{helder+09}.
More recently it has been pointed out that such anomalous line widths may be produced also by the dynamical backreaction of the so-called \emph{neutral return flux} induced by the population of hot neutrals produced via charge exchange immediately behind the shock \cite{neutri1}.
Such a neutral return flux leads to the formation of a neutral-induced precursor, in which the incoming fluid is slowed down and significantly heated up; 
hence, the shock becomes much weaker (i.e., with a reduced $r$) than it would be in an ionized medium and accelerates  particles with steeper spectra, even when the CR backreaction is included \cite{neutri3}.
Such a neutral feedback is expected to be important at SNR shocks when $\vsh\lesssim 3000$\,km\,s$^{-1}$, since for larger velocities ionization dominates over charge exchange and the neutral return flux vanishes.

\textbf{Other explanations.}
It also possible that some of the assumptions of the DSA theory are violated, especially at quasi-perpendicular shocks and for large shock velocities $\vsh\gtrsim 10^4$\,km\,s$^{-1}$.   
If the magnetic field is not uniform, but has a stochastic or braided structure, the transport of charged particles across the average direction of the field may be more complicated than simple pitch-angle scattering, resulting in anisotropic and/or inhomogeneous CR distributions \cite{kirk+96}.
Another possible reason for spectral steepening at quasi-perpendicular shocks is that the magnetic field may effectively sweep CRs through the shock, making their return to the shock for further acceleration less probable  \cite{bell+11}.
These scenarios assume that CR injection may happen also at quasi-perpendicular shocks, which is not granted in the absence of energetic seeds, though (\S\ref{sec:inj}).
Finally, it has been suggested that strong ion-neutral damping may steepen the CR spectrum (typically to $E^{-3}$) because of the partial evanescence of Alfv\'en waves above a critical energy of a few GeV \cite{malkov+12}.

\section{From accelerated particles to cosmic rays\label{sec:prop}}

To connect the CR spectra inferred in SNRs and the flux of CRs measured at Earth it is crucial to understand how accelerated particles are released into the ISM, which may happen either by escaping from upstream \cite{escape} or by joining the Galactic pool when the SNR fades away.
Trapped particles undergo strong adiabatic losses, so that only escaping particles can account for knee-like energies.
The spectrum of escaping particles is in principle different from that responsbile for the SNR emission, but it may still be a universal power-law $\propto E^{-2}$ as a result of the self-similar SNR evolution in the Sedov stage \cite{pz05,crspectrum}.
The presence of escaping particles is also revealed by the $\gamma$-ray emission from nearby MCs, which in principle contains interesting information about CR diffusion close to their sources (e.g., \cite{cs10,gabici+09,ohira+11}) and about CR propagation in weakly-ionized environments (e.g., \cite{morlino+15} and references therein).

When such a dual modality of particle release is accounted for, the total CR spectrum should be a bit steeper than that at the beginning of the Sedov stage \cite{efficiency}, consistent with the inferred values of $\delta\lesssim 0.5$ for the energy scaling of the Galactic confinement time.
$\delta\approx 1/3$ is also found to provide a more universal connection between the injected and the diffuse CR spectra, in the sense that for $\delta\gtrsim 0.5$ the intrinsic fluctuations in the distribution of SN explosions may significantly loosen the constraint $\alpha+\delta\approx 2.65$ and make the spectrum observed at Earth merely the result of a random realization \cite{ba12a}.
In addition, only relatively small values of $\delta$ can account for the low level ($\lesssim 1\%$) of CR anisotropy below the knee \cite{ba12b}.
Finally, $\delta\approx 1/3$ is also consistent with the steep CR spectra  ($\alpha\approx2.65-\delta\approx 2.3$) inferred from $\gamma$-ray observations of young SNRs (see \S\ref{sec:steep}). 

\subsection{CR self-confinement around sources}
The local excess of energetic particles immediately outside the sources should generate gradients in the CR distribution, which are expected to drive plasma instabilities that eventually tend to self-confine escaping particles.
Such a ``sphere of influence'' (the region where the contribution by one source dominates over the Galactic CR sea) becomes increasingly large at high energies because of the steepness of the diffuse spectrum and can be as large as a kpc.
The self-generated diffusion coefficient inferred in SNRs, $D_{\rm B}(E)\simeq 10^{20} E_{\rm GeV}/B_{100\mu {\rm G}}$, is about seven orders of magnitude smaller than the Galactic one at GeV energies, and none of these diffusion coefficients can properly parameterize the transport of CRs around their sources.
The strong non-linearity of the problem and the large scales involved make it very hard to address such a problem either analytically or numerically (see, e.g., \cite{bell+13, malkov15} for some attempts), especially because the saturation of the NRH instability in this context has not been assessed self-consistently, yet.

\subsection{CR self-confinement in the Galaxy}
On larger scales, the CR self-confinement is expected to be important also for determining their transport in the Milky Way, and in particular the inferred Galactic diffusion coefficient and the level of anisotropy in their arrival directions. 
By considering the non-linear resonant coupling between CRs and the spectrum of the ISM magnetic turbulence it is possible to work out a self-consistent solution for the CR equilibrium distribution and for the self-generated diffusion coefficient in the Galaxy \cite{blasi+12, aloisio+15}.
By assuming resonant streaming instability and non-linear Landau damping, three regimes of CRs diffusion can be individuated:
i) for $\gtrsim 200\,$GV CRs diffuse in the pre-existing waves with a Kolmogorov power spectrum  and $D\propto E^{1/3}$;
ii) below a few GV, CRs efficiently drive Alf\'enic modes and are advected with the self-generated waves at $v\approx v_A$;
iii) for intermediate energies CRs diffuse in the self-generated Alfv\'enic turbulence and $D\propto E^{1/2}$.
Such scalings are remarkably consistent with the ``spectral hardening'' observed at $\sim200\,$GeV (see \S\ref{sec:spectrum}).

The coupling of low-energy CRs with the ISM is hence predicted to be strong, which implies that CRs can ablate ionized gas from the Galactic disk (and possibly quasi-neutral material if charge-exchange is rapid enough) while they escape  the Galaxy \cite{ipavich75,breitschwerdt+91,ptuskin+97,recchia+15}.
Such CR-driven winds may play a pivotal role in galaxy formation, suppressing star formation and enriching galactic halos in heavy elements.

\section{Conclusions}
After many decades, the SNR paradigm still represents the most plausible explanation for the origin of Galactic CRs.
Thanks to the recent developments of $\gamma$-ray astronomy and kinetic simulations, observers and theorists have unraveled few long-standing issues, finding the direct evidence of hadronic acceleration in SNRs (\S\ref{sec:gamma}) and reproducing DSA ab initio (\S\ref{sec:DSA}), which also allowed to build a self-consistent model for ion injection (\S\ref{sec:inj}).
Nevertheless, there are still some tiles that need to be placed in the mosaic of understanding CR acceleration and transport.

The NRH instability is very likely the main channel of magnetic field amplification in SNRs and has the potential to foster the acceleration of PeV protons in very young sources, but a coherent description of its non-linear regime has not been put forward, yet. 
In addition, it is not clear whether the field configuration at its saturation is conducive to CR spectra steeper than the DSA prediction, either because of the finite velocity of the effective scattering centers or because of an incomplete confinement of energetic particles in its filamentary structures.

Despite recent progresses (\S\ref{sec:prop}), the CR-driven instabilities and damping mechanisms needed to describe CR transport around sources and in the Milky Way have not been singled out, yet.
Examples of effects that may play a pivotal role are ion-neutral damping and anisotropic transport, induced either by the large-scale Galactic magnetic field or by the anisotropic wave damping \cite{gs95}. 

The chemical composition of Galactic CRs is significantly much heavier than solar, especially in the knee region as a consequence of the slightly flatter spectra of species other than H. 
Such an enrichment may be due to dust sputtering (e.g., \cite{meyer+97}) or might as well be a signature that a significant fraction of the Galactic CRs is produced in high-metallicity star-forming regions.
If this were the case, a crucial question is whether collective effects (e.g., multiple SN explosions) lead to phenomena qualitatively different from the linear superposition of ``ordinary'' SNRs.

PIC simulations have just started unraveling injection, acceleration, and thermalization of electrons at non-relativistic shocks from first principles (\S\ref{sec:PIC}) but covering the multi-dimensional parameter space relevant for SNR shocks will require much more effort, especially for realistic mass ratios and 2D/3D setups. 
Yet, the prospects for studying  the physical processes crucial to the electron dynamics from first principles and extrapolate them to astrophysical scales are strong.

Understanding how SNRs affect their circumstellar medium is of primary importance for assessing the mechanisms that regulate star formation and, in turn galaxy, formation and evolution. 
CRs may play a crucial role in transporting and depositing energy and momentum in the ISM, and possibly also in the intracluster medium in galaxy clusters.
Now that the theory of CR acceleration at non-relativistic shocks has finally entered its quantitative age, the time may be ripe for embedding theoretically and observationally motivated sub-grid models in cosmological and galactic simulations, whose resolution is rapidly approaching the scales of individual SNRs.

\vspace{0.1cm}
\noindent {\bf Acknowledgments}\\
\noindent I would like to thank the ICRC organizers for their kind invitation to talk about such a pivotal topic, the University of Delaware and the IUPAP for the honor of receiving the Shakti P.\ Duggal Award 2015, and  all my collaborators, among which: E. Amato, G. Morlino, J. Park, T. Jones, H. Kang, M. Vietri, and in particular  P. Blasi and A. Spitkovsky. 
This research was partially supported by NASA (grant NNX14AQ34G to D.C.) and by the Max-Planck/Princeton Center for Plasma Physics.

\bibliography{Total}
\bibliographystyle{JHEP}

\end{document}